\documentclass[zpreprint,zbstnp]{zeus_paper}
\usepackage[american]{babel}
\newcommand{\Zdetdesc}{%
A detailed description of the ZEUS detector can be found 
elsewhere~\cite{zeus:1993:bluebook}. A brief outline of the 
components that are most relevant for this analysis is given
below.\xspace}
\chardef\usc=95
\chardef\til=126
\catcode`\@=11 % @ signs are now treated as letters
\DeclareRobustCommand\xdotspace{\futurelet\@let@token\@xdotspace}
\def\@xdotspace{%
  \ifx\@let@token.\else
  \ifx\@let@token\bgroup.\else
  \ifx\@let@token\egroup.\else
  \ifx\@let@token\/.\else
  \ifx\@let@token\ .\else
  \ifx\@let@token~.\else
  \ifx\@let@token!.\else
  \ifx\@let@token,.\else
  \ifx\@let@token:.\else
  \ifx\@let@token;.\else
  \ifx\@let@token?.\else
  \ifx\@let@token/.\else
  \ifx\@let@token'.\else
  \ifx\@let@token).\else
  \ifx\@let@token-.\else
  \ifx\@let@token\@xobeysp.\else
  \ifx\@let@token\space.\else
  \ifx\@let@token\@sptoken.\else
   .\space
   \fi\fi\fi\fi\fi\fi\fi\fi\fi\fi\fi\fi\fi\fi\fi\fi\fi\fi}
\catcode`\@=12 % @ signs are no longer letters

\newcommand{\stru}[2]{%
   \relax\ifmmode\hbox{\vrule height#1 depth#2 width0pt}%
   \else\vrule height#1 depth#2 width0pt\fi}
\newcommand{\Ronum}[1]{\uppercase\expandafter{\romannumeral#1}}
\newcommand{\ronum}[1]{\expandafter{\romannumeral#1}}
\DeclareRobustCommand{\LaTeXZ}{%
  \LaTeX\kern-.05em4\kern-.1em
  {\raisebox{-0.2ex}{$\scriptstyle\text{ZEUS}$}}\xspace}
\DeclareMathAlphabet{\mathbf}{OT1}{cmr}{bx}{sl}
\newcommand{\eVdist}{\kern-0.06667em}

\newcommand{\Gev}{{\text{Ge}\eVdist\text{V\/}}}

\newcommand{\mev}{{\,\text{Me}\eVdist\text{V\/}}}
\newcommand{\gev}{{\,\text{Ge}\eVdist\text{V\/}}}

\newcommand{\mrad}{\,\text{mrad}}

\newcommand{\slashfrac}[2]{%
  \raisebox{0.5ex}{\ensuremath #1}\kern-0.12em/\kern-0.08em
  \raisebox{-.8ex}{\ensuremath #2}}
\newcommand{\sqr}[3]{%
    {\vcenter{\hrule height.#3ex\hbox{\vrule width.#2ex height#1ex
     \kern#1ex\vrule width.#3ex}\hrule height.#2ex}}}

\catcode`\@=11 % @ signs are now treated as letters
\newcommand{\parenbar}{\mathpalette\p@renb@r}
\def\p@renb@r#1#2{\vbox{%
  \ifx#1\scriptscriptstyle \dimen@.7em\dimen@ii.2em\else
  \ifx#1\scriptstyle \dimen@.8em\dimen@ii.25em\else
  \dimen@1em\dimen@ii.4em\fi\fi \offinterlineskip
  \ialign{\hfill##\hfill\cr
    \vbox{\hrule width\dimen@ii}\cr
    \noalign{\vskip-.3ex}%
    \hbox to\dimen@{$\mathchar300\hfil\mathchar301$}\cr
    \noalign{\vskip-.3ex}%
    $#1#2$\cr}}}
\catcode`\@=12 % @ signs are no longer letters

\newcommand{\IP}{{\rm I$\kern-0.01667em$P}\xspace}

\mathchardef\qsm=63
\mathchardef\pls=43
\mathchardef\mns=512
\mathchardef\plm=518
\mathchardef\eql=61
\mathchardef\smallleft=300
\mathchardef\smallright=301
\mathchardef\les=316
\mathchardef\gre=318
\mathchardef\leq=532
\mathchardef\grq=533
\catcode`\@=11 % @ signs are now treated as letters
\newcounter{pict@width}
\newcounter{pict@height}
\newlength{\pict@scale}
\newcommand{\psfigadd}[4]{%
\setcounter{pict@width}{1*\ratio{#2+\pict@scale/2}{\pict@scale}}
\setcounter{pict@height}{1*\ratio{#3+\pict@scale/2}{\pict@scale}}
\setlength{\unitlength}{\pict@scale}
\hbox to #2{\hspace{-\fill}\begin{picture}(\thepict@width,\thepict@height)
\put(0,0){\psfig{figure=#1,width=#2,height=#3,clip=}}
\SetScale{0.283466457}
\SetWidth{1.763889}
{#4}
\end{picture}}
}
\newcounter{pict@widthfst}
\newcounter{pict@widthscd}
\newcounter{pict@widthtot}
\newcommand{\psfigaddtwo}[7]{%
\setcounter{pict@widthfst}{1*\ratio{#2+\pict@scale/2}{\pict@scale}}
\setcounter{pict@widthscd}{1*\ratio{#2+#4+\pict@scale/2}{\pict@scale}}
\setcounter{pict@widthtot}{1*\ratio{#2+#4+#6+\pict@scale/2}{\pict@scale}}
\setcounter{pict@height}{1*\ratio{#3+\pict@scale/2}{\pict@scale}}
\setlength{\unitlength}{\pict@scale}
\hbox{\hspace{-\fill}\begin{picture}(\thepict@widthtot,\thepict@height)
\put(0,0){\psfig{figure=#1,width=#2,height=#3,clip=}}
\put(\thepict@widthscd,0){\psfig{figure=#5,width=#6,height=#3,clip=}}
\SetScale{0.283466457}
\SetWidth{1.763889}
{#7}
\end{picture}}
}
\newcommand{\psfigror}[4]{%
\setcounter{pict@width}{1*\ratio{#2+\pict@scale/2}{\pict@scale}}
\setcounter{pict@height}{1*\ratio{#3+\pict@scale/2}{\pict@scale}}
\setlength{\unitlength}{\pict@scale}
\hbox{\begin{picture}(\thepict@width,\thepict@height)
\put(0,\thepict@height){\psfig{figure=#1,width=#3,height=#2,clip=,angle=270}}
\SetScale{0.283466457}
\SetWidth{1.763889}
{#4}
\end{picture}}
}
\newcommand{\psfigrol}[4]{%
\setcounter{pict@width}{1*\ratio{#2+\pict@scale/2}{\pict@scale}}
\setcounter{pict@height}{1*\ratio{#3+\pict@scale/2}{\pict@scale}}
\setlength{\unitlength}{\pict@scale}
\hbox{\begin{picture}(\thepict@width,\thepict@height)
\put(0,0){\psfig{figure=#1,width=#3,height=#2,clip=,angle=90}}
\SetScale{0.283466457}
\SetWidth{1.763889}
{#4}
\end{picture}}
}
\catcode`\@=12 % @ signs are no longer letters
\newlength\listtextwidth

\catcode`\@=11 % @ signs are now treated as letters
\newlength{\@tabfninsert}
\newlength{\@tabfnwidth}
\newcommand{\tabfootnote}[2]{%
  \setlength{\@tabfninsert}{0.8em}
  \setlength{\@tabfnwidth}{\textwidth}
  \addtolength{\@tabfnwidth}{-\@tabfninsert}
  \addtolength{\@tabfnwidth}{-0.4em}
  \noindent\makebox[\@tabfninsert][r]{\footnotesize$^{#1}$\hfil}\hfill%
  \parbox[t]{\@tabfnwidth}{\footnotesize #2\hfill}}
\catcode`\@=12 % @ signs are no longer letters
\def\citeCTD{{\cite{%
nim:a279:290,*npps:b32:181,*nim:a338:254%
}}\xspace}
\def\citeCAL{{\cite{%
nim:a309:77,*nim:a309:101,*nim:a321:356,*nim:a336:23%
}}\xspace}

\begin{document}
%------------------------------------------------------------------------------
%       Title sheet
%------------------------------------------------------------------------------
\title{
Exclusive electroproduction of $\phi$ mesons at HERA
}                                                       
                    
\author{ZEUS Collaboration}
%\draftversion{Submitted version - DESY report}
%\date{7 \ March 2005}

%\confname{International Conference on High Energy Physics}
%\confplacedate{16-22 August 2004, Beijing, China}
%\confabsnum{6-0248}
%\confsession{QCDsoft}
\prepnum{DESY 05-038}
\abstract{
 Exclusive electroproduction of $\phi$ mesons has been
studied in $e^\pm p$ collisions at $\sqrt{s}=318 \gev$
with the ZEUS detector at HERA 
using an integrated luminosity of $65.1$~pb$^{-1}$. 
The $\gamma^*p$ cross section is presented 
in the 
kinematic range  $2<Q^2<70 \gev^2$, $35<W<145 \gev$ and ${|t|}<0.6 \gev^2$.
The cross sections
as functions of $Q^2$, $W$, $t$ and helicity angle
$\theta_h$
are compared to cross sections for other vector mesons.  The ratios
$R$ of the cross sections for longitudinally and transversely
polarized virtual photons are presented as functions of $Q^2$ and $W$.
The data are also compared to predictions from theoretical models.
}

\makezeustitle

%------------------------------------------------------------------------------
%       Author list
%------------------------------------------------------------------------------
%===================================================================                               
%                                                                                                  
%  MEMBER NAME  AUTH124 (ZEUS)     M  TEX                                                          
%                                                                                                  
%  JH.: transformed to a format, which is suited as input for                                      
%       CONVERT, which automatically creates author-indices                                        
%                                                                                                  
%  Don't remove lines starting with a percent sign %,                                              
%  CONVERT may need them urgently !                                                                
%                                                                                                  
%=====================================================================                             

\def\3{\ss}
\pagenumbering{Roman}
\begin{center}                                                                                     
{                      \Large  The ZEUS Collaboration              }                               
\end{center}                                                                                       
  S.~Chekanov,                                                                                     
  M.~Derrick,                                                                                      
  S.~Magill,                                                                                       
  S.~Miglioranzi$^{   1}$,                                                                         
  B.~Musgrave,                                                                                     
  \mbox{J.~Repond},                                                                                
  R.~Yoshida\\                                                                                     
 {\it Argonne National Laboratory, Argonne, Illinois 60439-4815}, USA~$^{n}$                       
\par \filbreak                                                                                     
  M.C.K.~Mattingly \\                                                                              
 {\it Andrews University, Berrien Springs, Michigan 49104-0380}, USA                               
\par \filbreak                                                                                     
  N.~Pavel, A.G.~Yag\"ues Molina \\                                                                
  {\it Institut f\"ur Physik der Humboldt-Universit\"at zu Berlin,                                 
           Berlin, Germany}
% ...
\newpage                                                                                           
$^{\    1}$ also affiliated with University College London, UK \\                                  
% ...
                                                           %                                       
                                                           %                                       
% \par         % if index listing & table fit to 1 page, put gap here                              
\newpage   % alternatively: go to newpage, if page is too small                                    
                                                           %                                       
% \institute_references_start    % do not touch or move this line !                                
                                                           %                                       
\begin{tabular}[h]{rp{14cm}}                                                                       
$^{a}$ &  supported by the Natural Sciences and Engineering Research Council of Canada (NSERC) \\  
$^{b}$ &  supported by the German Federal Ministry for Education and Research (BMBF), under        
          contract numbers HZ1GUA 2, HZ1GUB 0, HZ1PDA 5, HZ1VFA 5\\                                
$^{c}$ &  supported in part by the MINERVA Gesellschaft f\"ur Forschung GmbH, the Israel Science   
          Foundation (grant no. 293/02-11.2), the U.S.-Israel Binational Science Foundation and    
          the Benozyio Center for High Energy Physics\\                                            
$^{d}$ &  supported by the German-Israeli Foundation and the Israel Science Foundation\\           
$^{e}$ &  supported by the Italian National Institute for Nuclear Physics (INFN) \\                
$^{f}$ &  supported by the Japanese Ministry of Education, Culture, Sports, Science and Technology 
          (MEXT) and its grants for Scientific Research\\                                          
$^{g}$ &  supported by the Korean Ministry of Education and Korea Science and Engineering          
          Foundation\\                                                                             
$^{h}$ &  supported by the Netherlands Foundation for Research on Matter (FOM)\\                   
$^{i}$ &  supported by the Polish State Committee for Scientific Research, grant no.               
          620/E-77/SPB/DESY/P-03/DZ 117/2003-2005 and grant no. 1P03B07427/2004-2006\\             
$^{j}$ &  partially supported by the German Federal Ministry for Education and Research (BMBF)\\   
$^{k}$ &  supported by RF Presidential grant N 1685.2003.2 for the leading scientific schools and  
          by the Russian Ministry of Education and Science through its grant for Scientific        
          Research on High Energy Physics\\                                                        
$^{l}$ &  supported by the Spanish Ministry of Education and Science through funds provided by     
          CICYT\\                                                                                  
$^{m}$ &  supported by the Particle Physics and Astronomy Research Council, UK\\                   
$^{n}$ &  supported by the US Department of Energy\\                                               
$^{o}$ &  supported by the US National Science Foundation\\                                        
$^{p}$ &  supported by the Polish Ministry of Scientific Research and Information Technology,      
          grant no. 112/E-356/SPUB/DESY/P-03/DZ 116/2003-2005 and 1 P03B 065 27\\                  
$^{q}$ &  supported by FNRS and its associated funds (IISN and FRIA) and by an Inter-University    
          Attraction Poles Programme subsidised by the Belgian Federal Science Policy Office\\     
\end{tabular}                                                                                      
                                                           %                                       
% \institute_references_end     % do not touch or move this line !                                 

%------------------------------------------------------------------------------
%       Text
%------------------------------------------------------------------------------
\pagenumbering{arabic} 
\pagestyle{plain}
\newcommand{\pom}{{I\hspace{-1mm}P}}

\section{Introduction}
\label{intro}

Exclusive electroproduction of vector mesons is a process 
which can be used to confront model predictions in a kinematic
region that includes the transition between soft and hard 
dynamics. The process also lends itself to detailed 
experimental investigation; the decay products of the 
vector meson can be precisely measured so that distributions in 
several variables can be studied with high resolution over a 
large phase space.

%%% COMMENTED OUT BY M. WING
%%% 
%%% Exclusive electroproduction of vector mesons
%%% is a process which lends
%%% itself to detailed experimental investigation.  The 
%%% decay products of the vector meson
%%% are precisely measured and distributions in many variables
%%% can be studied 
%%% with high resolution over a large phase space.
%%% Data from this process can be used to challenge model 
%%% predictions in a
%%% region that includes the transition between soft and hard dynamics.

Many measurements of exclusive production of 
%%($\rho,\omega$,$\phi$,$J/\psi,\psi^{\prime},\Upsilon$) 
vector mesons, $e\: p\rightarrow e\: V\: p$, have been made 
at HERA~\cite{
epj:c2:247, % rho - photo
epj:c6:603, % rho 96
epj:c12:393, % zeus - helicity rho
epj:c13:371,% h1-rho 95-95
pl:b539:25, % h1-nschc
zfp:c73:73, % omegaphoto
pl:b487:273, % omegadis
pl:b377:259, % phi- photo
pl:b380:220, % phi94
pl:b483:360, % h1 - phi
epj:c10:373, % h1-psi 96
epj:c24:345, % jpsiphoto
Chekanov:2004mw, %DIS J/psi
pl:b541:251, % h1 - psi(2s)
pl:b437:432, % zeus upsilon
pl:b483:23  % h1 - psi+upsilon
}.
 In this paper, results from a study of exclusive $\phi$ 
production are presented. The data sample represents a factor 
$\sim30$ increase over previously published HERA 
results~\cite{pl:b380:220,pl:b483:360}. The measurements were made for 
virtuality of the exchanged photon, $Q^2$, in the range 
$2<Q^2<70 \gev^2$ and for the photon-proton center-of-mass 
energy, $W$, in the range $35<W<145 \gev$. The cross sections are 
presented as functions of $Q^2$, $W$, the squared four-momentum 
transfer at the proton vertex, $t$, and helicity angle $\theta_h$. 
The $W$ dependence was also extracted in bins of $t$, and the 
ratio, $R$, of the cross sections for longitudinally and 
transversely polarized virtual photons, determined from the 
angular distribution of the decay products of the $\phi$ mesons, 
is presented as functions of $W$ and $Q^2$.

\section{Phenomenology}
\label{models}

 A simple picture for the process $e\:p \rightarrow e \:V \: p$, where
$V$ represents a vector meson, 
can be formulated in the rest frame of the
proton. In this frame, the virtual photon emitted from the
electron fluctuates into a quark-antiquark dipole, which then scatters
elastically off the proton. Long after the interaction, the $q\bar{q}$
pair forms a vector meson.
It can be shown that, at high energy,
the $q\bar{q}$ creation, scattering, and vector meson formation
are well separated in time so that
the cross section for the process can be factorized
into terms representing the $q\bar{q}$ coupling to the photon, 
the dipole scattering
cross section on the proton, and the final-state 
formation~\cite{pl:b324.knnz,pr:d50:3134}.
The dipole scattering cross section on the proton depends on the
transverse separation of the $q\bar{q}$ pair in the dipole.
The interactions of large dipoles are thought to be primarily
`soft' and therefore
described by Regge phenomenology \cite{collins:1977:regge}.  
On the other hand, the interactions of dipoles with small
transverse separation of the $q\bar{q}$ pair are expected to be `hard',
and therefore calculable in perturbative QCD (pQCD). 
In lowest-order pQCD, the process proceeds via the exchange of
two gluons between the proton and the $q\bar{q}$ dipole so that
vector-meson production is sensitive to the
gluon density in the proton~\cite{sovjnp:52:529}. 

 The transverse size
of the dipole is related to the transverse-energy scale of
the interaction, and can depend on $Q$ as well as on 
the quark mass, $M$~\cite{zfp:c57:89}.  
Large transverse-energy
scales preferentially select small dipole sizes. 
The impact parameter for the dipole scattering on the proton
depends on $\sqrt{|t|}$. Large $|t|$ preferentially selects small impact 
parameters.  The $|t|$ values in this analysis are small and no 
pQCD predictions are possible for the $|t|$ dependence in 
this regime.  The pQCD predictions for the $W$ and $Q^2$ dependence 
of the cross section should be more accurate as either or both of 
$Q$ or $M$ become large. 

%The precise value of the scale 
%and the
%appropriate combination of variables 
%where the transition to a regime where pQCD predictions are accurate is
%not known, and is an interesting topic of research.

The pQCD predictions depend on the square of the gluon density, leading to
a steep dependence of the $\gamma^*p$ cross section on $W$, 
in contrast to expectations from Regge phenomenology.
For example, a gluon density varying as $xg(x)\propto x^{-0.2}$,
where $x$ is the Bjorken scaling variable, would lead to
$\sigma(\gamma^* p \rightarrow V\: p) \propto W^{0.8}$.  
The Regge phenomenology 
expectation is $\sigma(\gamma^* p \rightarrow V \: p) \propto W^{0.2}$.
The energy dependence of the $\gamma^* p$ cross section is therefore a good
indicator of whether gluon exchange is dominant.
Data from exclusive $\rho$ 
production~\cite{epj:c6:603,epj:c13:371,pl:b539:25} 
show that the cross section $\sigma(\gamma^* p \rightarrow \rho\: p)$ 
rises with $W$ as $W^{\delta}$, where $\delta$ increases with $Q^2$
from about $0.2$ at $Q^2=0$ (photoproduction)
to about $0.7$ at $Q^2 \approx 30 \gev^2$.
% Only at high $Q^2$ is the energy dependence steep.
However, in the case of exclusive $J/\psi$ production
the cross section rises steeply with $W$ even
for photoproduction~\cite{Chekanov:2004mw,epj:c24:345,pl:b483:23}.
%It is therefore interesting to investigate $\phi$ production
%since it represents an intermediate mass state. 
The investigation of
elastic $\phi$ production is of particular interest since 
the $\phi$ is an $s\bar{s}$ state, and the mass of the strange quark
is not negligible in pQCD calculations.  Experimentally, the 
extraction of the $\phi$ signal
is very clean due to the narrow width of the state.

%$M^2_{\phi}\approx 1$~GeV$^2$ and the transition from soft to hard
%interactions can be probed.

The measurement of the variation of the 
$W$ dependence of the cross section with $t$ 
also provides a good
test of the validity of pQCD calculations.  In pQCD, little variation is
expected, while for soft hadronic processes the $W$ dependence 
varies as $\sigma\propto W^{4(\alpha_{\pom}(t)-1)}$, with 
$\alpha_{\pom}(t)=1.08+0.25t$~\cite{pl:b348:213,pr:d10:160}. 

Specific predictions for $\phi$ meson production are available in the
dipole model realization (FS04) 
of Forshaw and Shaw~\cite{FS04}.  The predictions
depend on the strange-quark mass used in the photon wavefunction, on the
parametrization of the dipole-proton scattering cross section, and on the 
vector-meson wavefunction.  The gluon density does not appear explicitly
but is implicit in the dipole scattering cross section.
The predictions shown in this paper employed
a strange-quark mass of $200$~MeV, and a Gaussian wave-function 
for the $\phi$-meson~\cite{misc:forshaw:private}.  
The dipole-proton cross section is obtained from the
best fit to the total $\gamma^*p$ cross section~\cite{FS04} 
and implies a saturation of the cross section 
as $x$ decreases.  
%This model is labeled FS04.
%The normalization of the cross section is taken from the authors of this
%prediction.

Specific predictions for the $\phi$ meson are also
available in the model of Martin,
Ryskin and Teubner (MRT)~\cite{pr:d62:14022}.
In the MRT model, the properties of the event are determined
completely from the features of the photon wavefunction and the gluon density
in the proton. Parton-hadron duality is invoked to relate the parton-level
cross section to the cross section for the produced vector meson.
% The MRT
%model gives a prediction for $\frac{d\sigma}{dt}|_{t=0}$, and must therefore
%be normalized by $1/b$, where $b$ is the slope of the $t$-distribution, to get
%the total cross section.  A parametrization of the $b$ dependence found in this
%analysis was used for the normalization.

As NLO corrections are not fully taken into account, the model
calculations come with significant normalization uncertainties.

The data of this analysis are also compared to results 
from other vector mesons.
The results are presented as a function of $Q^2+M^2_V$, where
$M_V$ is the vector meson mass, for cross sections, and
as a function of $Q^2/M^2_V$ for the spin-density matrix element, 
$r_{00}^{04}$, to test for scaling in these variables.

A recent review of models and data can be consulted for more detailed 
information on vector meson production~\cite{ivanov:nikolaev:savin}.

% ----------------------------------------------------------------------------
%       Experimental set-up
% ----------------------------------------------------------------------------
\section{Experimental set-up}
\label{sec-exp}

%%% TAKEN FROM INTRODUCTION: M. WING
%%%
The data were collected during
1998-2000 with the ZEUS detector and correspond to an integrated
luminosity of $15.0$~pb$^{-1}$ for $e^{-}p$ and $50.1$~pb$^{-1}$ for
$e^{+}p$ collisions with a proton energy of $920\gev$ and an $e^{\pm}$ energy
of $27.5\gev$.
Since no dependence on the lepton charge is expected,
the two data sets were combined\footnote
{Hereafter, both $e^+$ and $e^-$ are referred to as electrons, 
unless explicitly stated otherwise.}.
\Zdetdesc

%%~\citeCTD
Charged particles were reconstructed in the central tracking detector 
(CTD)~\citeCTD covering the polar-angle{\footnote{
The ZEUS coordinate system is a right-handed Cartesian system, with the $Z$
axis pointing in the proton beam direction, referred to as the ``forward
direction'', and the $X$ axis pointing left towards the center of HERA.
The coordinate origin is at the nominal interaction point.\xspace}}
region $15^\circ<\theta<164^\circ$. The transverse-momentum resolution for
full-length tracks is $\sigma(p_T)/p_T=0.0058p_T\oplus0.0065\oplus0.0014/p_T$,
with $p_T$ in $\Gev$.

The high-resolution uranium calorimeter (CAL)~\citeCAL consists of three parts:
the forward (FCAL), the barrel (BCAL) and the rear (RCAL) calorimeters. 
Each part
is subdivided transversely into towers and longitudinally into an 
electromagnetic section (EMC) and either one (RCAL) or two (FCAL and BCAL) 
hadronic sections. The CAL covers $99.7\%$ of the total solid angle.
Under test-beam conditions, the CAL single-particle relative energy
resolution is $\sigma(E)/E=0.18/\sqrt{E}$ for electrons and 
$\sigma(E)/E=0.35/\sqrt{E}$ for hadrons, with $E$ in $\gev$.

The forward plug calorimeter (FPC)~\cite{nim:a450:235} was a lead-scintillator
sandwich calorimeter with readout via wavelength-shifter fibers.
It was installed in the beamhole of the FCAL 
and extended the pseudorapidity coverage of the
forward calorimeter from  $\eta \lesssim 4$ to $\eta \lesssim 5$.
It has since been removed to accomodate HERA magnets for the high-luminosity 
HERA II run.

The small-angle rear tracking detector (SRTD)~\cite{nim:a401:63} consists of
two planes of scintillator strips read out via optical fibers.
It is attached to the front face of the RCAL
and covers an angular range between $162^\circ < \theta < 176^\circ$. 
The SRTD provides a transverse position resolution for the scattered electron 
of 
0.3~cm~\cite{epj:c24:345} corresponding to an angular resolution of $2\mrad$.

 The hadron-electron separator installed in the RCAL (RHES) consists of silicon diodes 
placed at a longitudinal depth of three radiation lengths. The RHES provides an electron 
position resolution of 0.5~cm if at least two adjacent pads are hit by the 
shower \cite{nim:a277:176}.

The luminosity was determined from the rate of the bremsstrahlung process
$ep\rightarrow e\:\gamma\: p$, 
where the photon was measured with a lead-scintillator
calorimeter~\cite{acpp:b32:2025} placed in the HERA tunnel at \mbox{$Z= -107$~m} 
in the HERA tunnel.
% ----------------------------------------------------------------------------
%       Kinematics
% ----------------------------------------------------------------------------
\section{Kinematics and cross sections}
\label{sec-kinem}
The following kinematic variables are used
to describe exclusive $\phi$ production,
$$e(k)\: p(P)\rightarrow e(k^{\prime})\: \phi(v)\: p(P^{\prime})\;\;,$$
where $k$, $k^{\prime}$, $P$, $P^{\prime}$ and $v$ are, respectively,  
the four-momenta of the incident electron, scattered electron, 
incident proton, scattered proton and the $\phi$ meson:
\begin{itemize}
\item
$Q^2 = -q^2 = -(k - k^{ \prime})^2$, the negative four-momentum squared of
the virtual photon; 
\item
$W^2=(q+P)^2$, the squared invariant mass
of the photon-proton system; 
\item
$y=(P\cdot q)/(P\cdot k)$, the fraction of the 
electron
energy transferred to the proton in the proton rest frame;
\item 
$x=Q^2/(2 P\cdot q)$, the Bjorken variable;
\item
$t = (P-P^{\prime})^2$, 
the squared four-momentum transfer  at the proton vertex.
\end{itemize}

The kinematic variables were reconstructed with the ``constrained'' 
method~\cite{epj:c6:603} which uses the momentum vector of the  
$\phi$  and the polar and azimuthal angles of the scattered electron.

The $ep$  cross section can be expressed in terms
of the transverse, $\sigma_T$,  
and longitudinal,   
$\sigma_L$, virtual photoproduction cross sections as
\begin{equation*}
\frac{d^2\sigma^{ep\rightarrow e \phi p}}{dydQ^2} = 
\Gamma_T(y,Q^2) 
\left (  \sigma_T + \epsilon \sigma_L \right ),
\end{equation*}
where $\Gamma_T$ is the flux of transverse virtual photons~\cite{pr:129:1834}
and $\epsilon$ is the ratio of 
longitudinal and transverse virtual-photon fluxes, given by
$\epsilon = 2(1-y)/(1+(1-y)^2)$.
In the kinematic range studied here, $\epsilon$
lies in the range $0.975<\epsilon<1$, with an average value of $0.99$.

The virtual photon-proton cross section, 
$\sigma^{\gamma^* p \rightarrow \phi p} \equiv \sigma_T + \epsilon \sigma_L$,
can be used to evaluate the total exclusive cross section, 
$\sigma_{{\rm tot}}^{\gamma^* p \rightarrow \phi p} \equiv \sigma_T + \sigma_L$,
through the relation
\begin{equation*}
%\label{sigmatot}
\sigma_{\rm tot}^{\gamma^* p \rightarrow \phi p} = \frac{1+R}{1+\epsilon R}  \sigma^{\gamma^* p \rightarrow \phi p},
\end{equation*}
where $R=\sigma_L/ \sigma_T$
is the ratio of the cross sections for longitudinal and transverse
photons. 
The helicity structure of $\phi$ production is used to determine $R$ 
as described in Section~\ref{section-R}.

% ----------------------------------------------------------------------------
%       Event selection
% ----------------------------------------------------------------------------
\section{Reconstruction and selection of events}
\label{sec-eventsel}
The signature of exclusive $\phi$ electroproduction,
$ep\rightarrow e\: \phi\:p$,
consists of the scattered electron
and two oppositely charged kaons from the $\phi$ decay. 
The scattered proton is deflected through a small angle and escapes undetected
down the beampipe.  The data selection and analysis are described in
detail elsewhere~\cite{thesis:miro}.  A brief outline is given here.

The events were selected online by a three-level
trigger~\cite{trigger:1992,nim:a355:278}. The trigger required a
scattered electron in the CAL with energy greater than $7\gev$, a minimum of two 
and a maximum of five
tracks reconstructed with the CTD and less than $5 \gev$ in the towers of
the FCAL closest to the beampipe.
These cuts reduced the rate of background events while preserving 
high efficiency ($>99$~\%) for the signal events.
 
The following criteria were applied offline to reconstruct and select the 
events:
\begin{itemize}
\item 
the identification and energy measurement of the scattered electron
used information from the CAL. The energy was required to satisfy $E>10
\gev$. The impact point of the electron on the CAL
was measured using three detectors: SRTD, HES
and CAL. Given its superior position resolution, 
preference was given to the measurement from the SRTD. This was improved by
the position obtained from HES or CAL when applicable. To
ensure full containment of the electromagnetic shower and good
position reconstruction, fiducial cuts were applied to the impact
position of the electron on the face of the RCAL;
\item
the $\phi$ mesons were reconstructed from the properties of the
decay kaons.  No
particle identification was used. Events with
two tracks of opposite charge each
with $p_T>0.15\gev$ and $|\eta|<1.7$ were selected.  These tracks were 
assigned the kaon mass and the invariant mass was formed.  
Track combinations with invariant masses falling within an allowed
mass window were selected (see below).
Events with additional tracks not associated with the scattered electron
or with kaon decays were rejected; 

\item 
the position of the reconstructed vertex was required 
to be compatible with that of an $ep$ collision, $|Z_{\rm VTX}|<50$~cm.
The radial distance of the reconstructed vertex from the nominal
beamline was required to be less than $0.8$~cm 
to remove $K^0_S\rightarrow \pi^+\pi^-$ decays;
\item 
to remove events with large initial-state radiation, 
\mbox{$45<E-P_Z<65 \gev$} was imposed, where the longitudinal energy-momentum variable
$E-P_Z$ is calculated using the momenta of the two kaons and the 
scattered electron (the masses are neglected).  This variable is peaked
at twice the electron beam energy ($55$~GeV) for non-radiative DIS events.
The $E-P_Z>45$~GeV cut removed events with a radiated photon of more than
$5$~GeV;
\item 
to suppress non-exclusive events, the energy of each CAL cluster not
associated with either of the final-state kaons or the scattered electron
was required to be less than $0.3 \gev$. To suppress further the
contamination from proton-dissociative events, $e p\rightarrow e\: \phi\:
Y$, the energy in the FPC was required to be less than $1 \gev$.
These cuts restrict the mass of the proton-dissociative system, $Y$, to
$M_Y \lesssim 2.3 \gev$.
\end{itemize}

\begin{figure}[p]
\vfill
\begin{center}
\epsfig{file=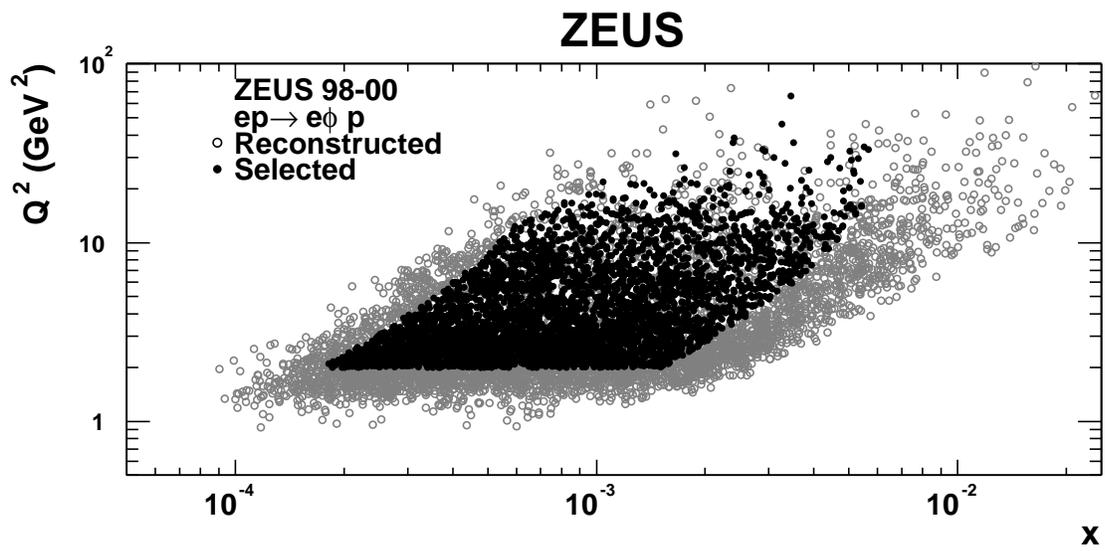,width=16.cm}
\end{center}
\caption{The distribution of the events in the kinematic plane of
Bjorken x and $Q^2$. The open grey dots represent all reconstructed events
while the solid black dots are events in the accepted kinematic range (Section~\ref{sec-eventsel}).}
\label{fig-scattq2x}
\vfill
\end{figure}

Events were required to be in a kinematic region where the properties of
the final-state
particles are well measured. Additionally, the kinematic range was limited
to the region where the acceptance
varies only slowly with the kinematic variables.  This led to the following 
selection:
\begin{eqnarray*}
Q^2 & > & 2\gev^2 \;\; , \\
|t| & < & 0.6\gev^2 \;\; ,\\
33.75 \gev +(1.25 \gev^{-1}\cdot Q^2) < & W & < 100\gev +(3.7\gev^{-1}\cdot Q^2)
\;\; ,
\end{eqnarray*}
with $Q^2$  given in GeV$^2$  in the last expression. The distribution of
selected events in the $x$-$Q^2$ plane for $1.01<m_{KK}<1.04\gev$ is
shown~in~Fig.~\ref{fig-scattq2x}.

% The number of events is obtained using an unbinned likelihood fit to
%the invariant mass spectrum for the $K^{+}K^{-}$ pairs in the range
%$2m_{K}<m_{KK}<1.1\gev$. A total of 4058 events were extracted in the accepted
%kinematical range and $1.01<m_{KK}<1.04\gev$.  The distribution of the
%events in the $x$-$Q^2$ plane is shown in Fig.~\ref{fig-scattq2x}.

% ----------------------------------------------------------------------------
%       Monte Carlo 
% ----------------------------------------------------------------------------
\section{Monte Carlo simulation}
\label{sec-mc}
The acceptance and the effects of the detector response were determined
using samples of Monte Carlo (MC) events. 
All generated events were passed through the standard ZEUS detector
simulation, based on the {\sc Geant}~3.13 program~\cite{tech:cern-dd-ee-84-1},
the ZEUS trigger simulation package, and the ZEUS reconstruction software.

The exclusive process $ep\rightarrow e\: \phi\:p$ 
was modelled using the {\sc Zeusvm}~\cite{thesis:muchorowski:1996} 
MC generator
interfaced to 
{\sc Heracles}~4.6.1~\cite{cpc:69:155-tmp-3cfb28c9,*spi:www:heracles}
%~\cite{spi:www:djangoh11,*spi:www:heracles,*cpc:69:155} 
to account for first-order QED radiative effects. 
%For the extraction of acceptances and other MC corrections, t
The parameters describing the 
$W, Q^2, t$ and $\theta_{h}$ distributions in the MC simulation were adjusted
such that the MC simulation reproduced the data distributions.  
%The $\gamma^*\;p \rightarrow \phi\;p$ cross section was parametrized as 
%$W^{\delta}e^{-b|t|}(M^2_{\phi}+Q^2)^{-n}$. 
%The parameter values $n=X$, $\delta=X$ 
%and $b=X\gev^{-2}$  were used in the event generator.
%S-channel helicity conservation (SCHC) was assumed in the event
%generation, 
%and the ratio of the cross sections for longitudinal and
%transverse photons was assumed to be $X \cdot (Q^2/M^2_{\phi})$.
%For the extraction of acceptances and other MC
%corrections, the MC events were reweighted using the parameters which gave
%the best description of the observed data.  
The typical acceptance for $\phi \rightarrow K^+K^-$ 
increases from $20$\% at $Q^2=2 \gev^2$ to $60$\% for $Q^2>10
\gev^2$.  The acceptance is small at the smaller $Q^2$ due to the fiducial
cuts on the impact point of the electron.

 Proton-dissociative events, $e p\rightarrow e\: \phi\: Y$, were modelled
using the generator {\sc
Epsoft}~\cite{thesis:adamczyk:1999,*thesis:kasprzak:1994}.  The
$\gamma^*\:p \rightarrow \phi\:Y$ cross section was parametrized as
\begin{equation}
\frac{d^2\sigma^{\gamma^* p \rightarrow \phi Y}}{dt~dM^2_Y}
\propto f(Q^2,W) e^{-b|t|}M^{-\beta}_Y
\label{eq-epsoft}
\end{equation}
with the same $Q^2$ and $W$ dependence, $f(Q^2,W)$,
 as for the exclusive case.  The values of $b$ and $\beta$ are discussed in
Section~\ref{pdiss_background}.

% ----------------------------------------------------------------------------
%       Mass spectra~\ref{fig-inv_mass}d and ~\ref{fig-inv_mass}e,
% ----------------------------------------------------------------------------
\section{Extraction of the $\phi$ signal}
\label{sec-mass}
\begin{figure}[p]
\vfill
\begin{center}
\epsfig{file=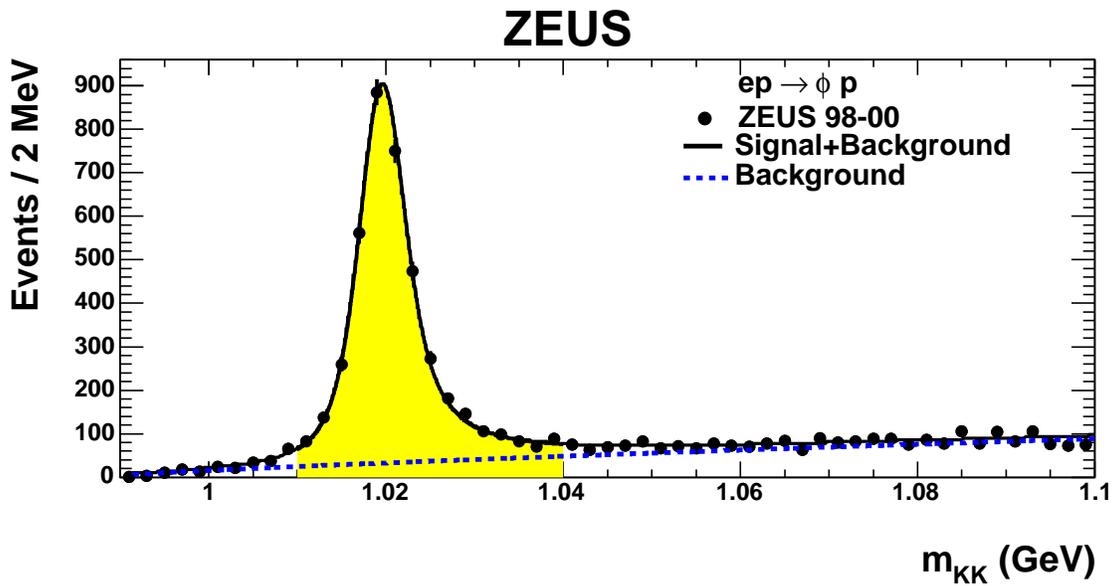,width=16.cm}
\end{center}
\caption{
Invariant mass distribution of the $K^+K^-$ candidate pairs.
The solid line shows the result of a fit including signal+background,
while the dashed line shows the background component only.  The shaded region
indicates the mass range used for cross-section calculations.
The error bars indicate the statistical uncertainties.
}
\label{fig-inv_mass}
\vfill
\end{figure}

Figure~\ref{fig-inv_mass} shows the invariant-mass distribution of the
$K^+K^-$ pairs obtained after the selection described in
Section~\ref{sec-eventsel}. A total of 3642 $\phi$ candidates 
were found after subtraction of non-resonant background in the mass
range $1.01<m_{KK}<1.04\gev$.  The natural width of the
resonance is comparable to the detector mass resolution, which is $1.8\mev$
for tracks at central rapidity, deteriorating to $\sim5\mev$ for
tracks crossing the CTD through large angles. This resolution is well
described by the MC simulation.
% The smooth non-resonant
%background is dominated by exclusive $\rho \rightarrow \pi^+\pi^-$
%production.

\subsection{Non-resonant background}
\label{sec-signal}
The non-resonant background was estimated for each bin in which a
cross section was extracted with an unbinned likelihood fit to the
invariant-mass distribution. The assumed functional form was a p-wave
relativistic Breit-Wigner convoluted with a Gaussian resolution
function for the signal plus a background function with the shape
$a(m_{KK}-2m_K)^b$. The background, estimated from the fit,
was subtracted from the number
of observed events in the mass window. The contribution of the
non-resonant background in the mass range of the signal 
is typically 18\% at
$Q^2=2.4 \gev^2$, decreasing to 5\% at $Q^2=13 \gev^2$.

\subsection{Proton-dissociative background}
\label{pdiss_background}
The remaining  source of background  consists of 
$\phi$ production accompanied by proton dissociation,
$e\: p\rightarrow e\:\phi\:Y$, 
where the particles from the breakup of the proton  
are not detected.  A similar study to that done for $J/\psi$ 
production~\cite{epj:c24:345} was performed.

Proton-dissociative events with observed proton breakup
were studied using a sample of diffractive 
events selected as described in Section~\ref{sec-eventsel}, 
with the following exceptions:
\begin{itemize}
\item the elasticity criteria (last bullet in  Section~\ref{sec-eventsel})
were not applied and the energy in FPC was required to satisfy
$E_{\rm FPC}>1\gev$;
\item a pseudorapidity gap $\Delta \eta>3$ was required between the energy
deposits from the system $Y$ and the $\phi$ decay products.
\end{itemize}
The sample of proton-dissociation events identified with 
the FPC contained $300$ $\phi$ 
candidates for $|t|<1 \gev^2$
in the kinematic range $45 <W< 135\gev$ and $Q^2>2\gev^2$.  The $W$,
$Q^2$ and $\theta_h$ dependences were found to be the same as for
the elastic data sample. The $t$ dependence was found to have a slope
(see Eq.~1) 
decreasing with $Q^2$ as $b=(5.1\pm1.3)/(1+R(Q^2))\gev^{-2}$ with $R$
given in Section~\ref{section-R}. The MC distribution of $M^2_Y$ was
tuned to describe the FPC energy distribution, yielding $\beta=2.0\pm
0.5$. 

The fraction of proton-dissociative events in the final sample,
averaged over $t$ for $|t|<0.6\gev^2$,
was $f_{\rm p-diss}=(7.0\pm 0.4^{+4.2}_{-2.8})\%$,
independent of $W$, $Q^2$ and $\theta_h$.
The uncertainty is dominated by the uncertainty in the modelling of the $M_Y$
spectrum for $M_Y<3 \gev$, and by the simulation of the proton-remnant final
state.  The uncertainty of the behavior of the $M_Y$ spectrum at small masses
was parametrized by allowing $\beta$ to vary in the range
$1.5 <\beta<3.0$.  
The uncertainties on the simulation of the
proton-remnant final state were estimated by a comparison of different MC
simulations.

The fraction of proton-dissociative events 
increases from 4.5\% for $0<|t|<0.08\gev^2$ 
to 14.5\% for $0.35<|t|<0.6\gev^2$.
The cross sections presented in the next sections 
were corrected for this background 
in bins of $t$, and globally in $W$, $Q^2$ and $\theta_h$.

% ----------------------------------------------------------------------------
%       xsection calculation
% ----------------------------------------------------------------------------
\section{Results}
\subsection{Cross sections}
\label{sec-xsec}
In each bin of a kinematic variable,
the $ep$ cross section was extracted 
using the formula
\begin{equation*}
\sigma^{ep\rightarrow e\:\phi\:p} = 
\frac{(N_{\rm data} -N_{\rm bgd})  (1 - f_{\rm p-diss} ) } {\cal A B  L},
\end{equation*}
where $N_{\rm data}$ is the number of events in the data
and $N_{\rm bgd}$ is the number 
of events from the non-resonant background.
The overall acceptance is denoted as ${\cal A}$,
${\cal B}$ accounts for the 
$\phi\rightarrow K^+K^-$ decay branching ratio
$(49.2\pm0.6)\%$~\cite{pr:d66:31}, 
and ${\cal L}$  is the integrated luminosity.
 
The total exclusive photon-proton cross section was calculated as
\begin{equation*}
\sigma_{\rm tot}^{\gamma^* p \rightarrow \phi p}(Q^2,W,t) 
= \sigma_{\rm tot,MC}^{\gamma^* p \rightarrow \phi p}(Q^2,W,t)
\cdot \frac{\sigma^{ep\rightarrow e\:\phi\:p}_{\rm DATA}}
{\sigma^{ep\rightarrow e\:\phi\:p}_{\rm MC}} \; .
\end{equation*}
In this way, the MC simulation 
was used to correct for radiative effects, detector
acceptance, cut efficiencies and the shape of the cross section within
the bin.
  
The cross sections were measured for $|t|<0.6\gev^2$.  The results
were then extrapolated to the full $t$ range 
assuming $d\sigma/dt \propto e^{-b|t|}$.
The correction factor needed to extrapolate to the cross section
integrated over $t$ was evaluated bin-by-bin
using the measured value of $b$ for the given $Q^2$, and ranged from 
$1.5$~\% to $5.0$~\%.  The cross 
sections are quoted for the mass range 
\mbox{$2m_K < m_{KK} < M_{\phi} + 5\Gamma$},
where $M_{\phi}=1019.4$~MeV and $\Gamma=4.458$~MeV~\cite{pr:d66:31}.

% ----------------------------------------------------------------------------
%       Systematics
% ----------------------------------------------------------------------------
\subsection{Systematic uncertainties}
\label{sec-syst}
The systematic uncertainties on the measured cross sections were determined by
varying the selection cuts and by modifying the analysis procedure. 
The sources of systematic uncertainties considered~\cite{thesis:miro} 
were similar to those of
previously published analyses~\cite{epj:c21:443,epj:c24:345}.
Except for the two lowest $Q^2$ bins in the
evaluation of $d \sigma/d Q^2$, the systematic uncertainties
 (excluding normalization uncertainty) were smaller than
the statistical errors.
\begin{itemize}
\item The electron position reconstruction is critical in determining the
acceptance and in the kinematic-variable 
reconstruction.  Possible misalignments of the SRTD and CAL were estimated to
be $2$~mm, and the effect of such misalignments
was tested via MC simulations.  In addition,
the electron fiducial cut was varied by $5$~mm. The incomplete knowledge
of the reconstructed electron position
generally provided the largest source of uncertainty.
\item The elasticity cut is important in suppressing non-exclusive events.
%Randomly triggered data events were used to estimate the effects of
%calorimeter noise and the noise was implemented in the MC simulation.
The energy distribution of the most energetic cluster not
assigned to the electron or one of the kaons was compared in data and
MC simulation.  The MC simulation was found to reproduce the data distributions
well, indicating a very small non-exclusive contribution with extra
particles in the central region. The cut was varied from $0.3$ to
$0.4 \gev$ in both data and MC simulation to determine the
uncertainties.
\item The stability of the fitting procedure and extraction of the signal
was checked by reducing the fit range to $m_{KK}<1.07\gev$ and changing the
accepted mass range to $1.01<m_{KK}<1.035\gev$,
respectively. Except for bins with a small number of events the
contribution to the overall uncertainty was small.

\item Uncertainties from 
the dependence on the MC parametrizations were also estimated and
found to be small compared to other sources of systematic
uncertainties.
%\item The effect of all cuts was checked by varying them within reasonable
%bounds. The effects were small.
\item The normalization uncertainty was different for different
measurements.  All measurements had a normalization uncertainty
due to uncertainties in the integrated luminosity, $\pm 2.5\%$, 
and due to the uncertainty in the decay branching
ratio, $\pm 1.2\%$, leading to a combined uncertainty of $\pm 2.8\%$. In
the case of the $d\sigma/d Q^2$ and $d\sigma/d W$ measurements, the
proton-dissociation background was subtracted globally and the normalization
uncertainty increased to $^{+4.1}_{-5.3}$~\%. 
The uncertainty due to the subtraction of proton-dissociative
background for the $d\sigma/ d t$ measurement resulted in a $t$-dependent
uncertainty as discussed in Section~\ref{pdiss_background}.  For the
angular distributions, the normalization uncertainty does not appear
as only the shapes of the distributions were measured.  
\end{itemize}

The total systematic uncertainty (excluding the normalization uncertainty)
was determined by adding the individual contributions in
quadrature.  The correlated and uncorrelated systematic uncertainties
were evaluated separately. The typical size of the systematic
uncertainty (excluding the normalization uncertainty) 
was $5$\% for the cross sections
in $W$ and $Q^2$ bins, and $5$\% at small $|t|$, increasing to $10$\%
at $|t|=0.6 \gev^2$.

% ----------------------------------------------------------------------------
%       W and Q2 dependence 
% ----------------------------------------------------------------------------
\subsection{Dependence on $W$ and $Q^2$}
\label{sec-wdep}
The  cross-section $\sigma_{\rm tot}^{\gamma^* p \rightarrow \phi p}$,
measured as a function of $W$ and $Q^2$, 
is shown in Fig.~\ref{fig-xsec}a and given in
Table~\ref{tab:xsecW}.  In addition to the measured cross section, this table
gives the
kinematic range over which the measurement was performed, the value of the
kinematic variables at which the cross section is quoted, the acceptance, and
the background-corrected number of $\phi$ events.  
The much higher statistics available for this analysis
allow measurements of double-differential cross sections.

The cross sections were fitted to a dependence
$\sigma \propto W^{\delta}$ with results presented in 
Table~\ref{tab:delta-q2}.  The measured values of $\delta$ 
show no $Q^2$ dependence within the present uncertainties.
The values of $\delta$  are
compared to those from previous H1~\cite{epj:c13:371} and 
ZEUS~\cite{epj:c2:247,epj:c6:603,epj:c24:345,Chekanov:2004mw} 
measurements
% for different vector mesons
in Fig.~\ref{fig-xsec}b, where the data are plotted as a function of
$Q^2+M_V^2$.  The values of $\delta$ scale with this variable 
within the present uncertainties.

The $Q^2$ dependence of the cross section, given in Table~\ref{tab:xsecQ2}, 
is shown in Fig.~\ref{fig-Q2xsec}a for $W=75 \gev$.
The data are compared 
to previous ZEUS~\cite{pl:b377:259,pl:b380:220} and 
H1~\cite{pl:b483:360} results.  The 
$\phi$ data from this analysis were fitted with a function of the form
$\sigma \propto (Q^2 + M_{\phi}^2)^{-n}$ expected in the 
VDM~\cite{prl:22:981}. 
The new high-precision data 
show that the $Q^2$ dependence of the cross section
cannot be fitted with a single value of $n$
over the $Q^2$ range of this analysis. 
The
fit parameter $n$ was found to vary from $n=2.087\pm0.055({\rm
stat.})\pm0.050({\rm syst.})$ for $2.4 \le Q^2 \le 9.2 \gev^2$ to
$n=2.75\pm0.13({\rm stat.})\pm0.07({\rm syst.})$ for
$9.2\le Q^2 \le 70 \gev^2$.
There are several possible causes that could lead to this behavior, including
a variation of $R$ with $Q^2$, 
the dependence of $\alpha_S$
on $Q^2$ and the changing $Q^2$ dependence of the gluon density at fixed $W$.
 The
longitudinal and transverse components of the cross section have been
separately extracted using the measured value of $R$ 
(see Section~\ref{section-R}), and are shown in Fig.~\ref{fig-Q2xsec}b.
The difference in the $Q^2$ dependence of the cross section for the two
helicity components is clearly seen.

\subsubsection{Comparison to Models}
\begin{figure}[p]
\vfill
\begin{center}
\epsfig{file=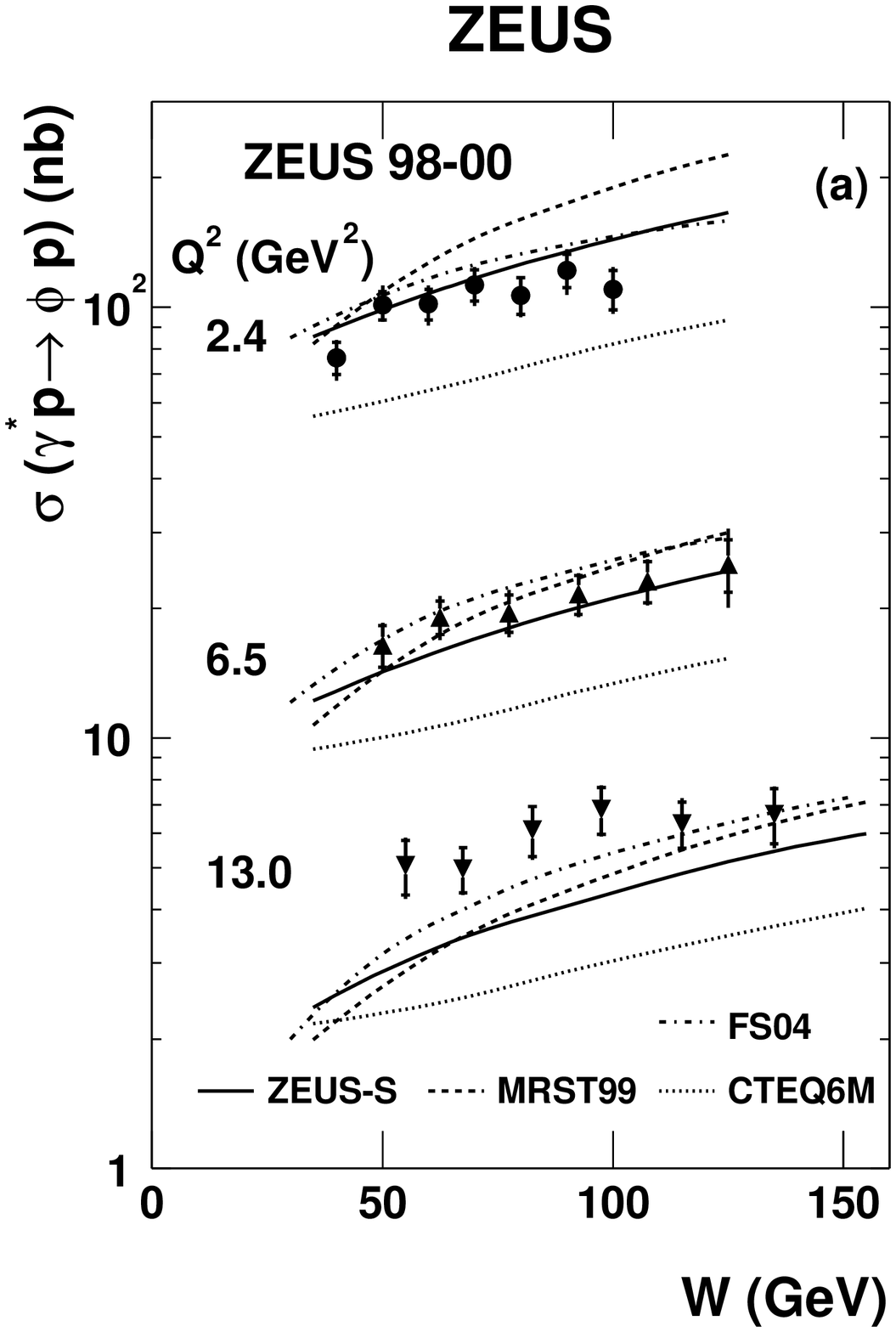,width=8.cm}\epsfig{file=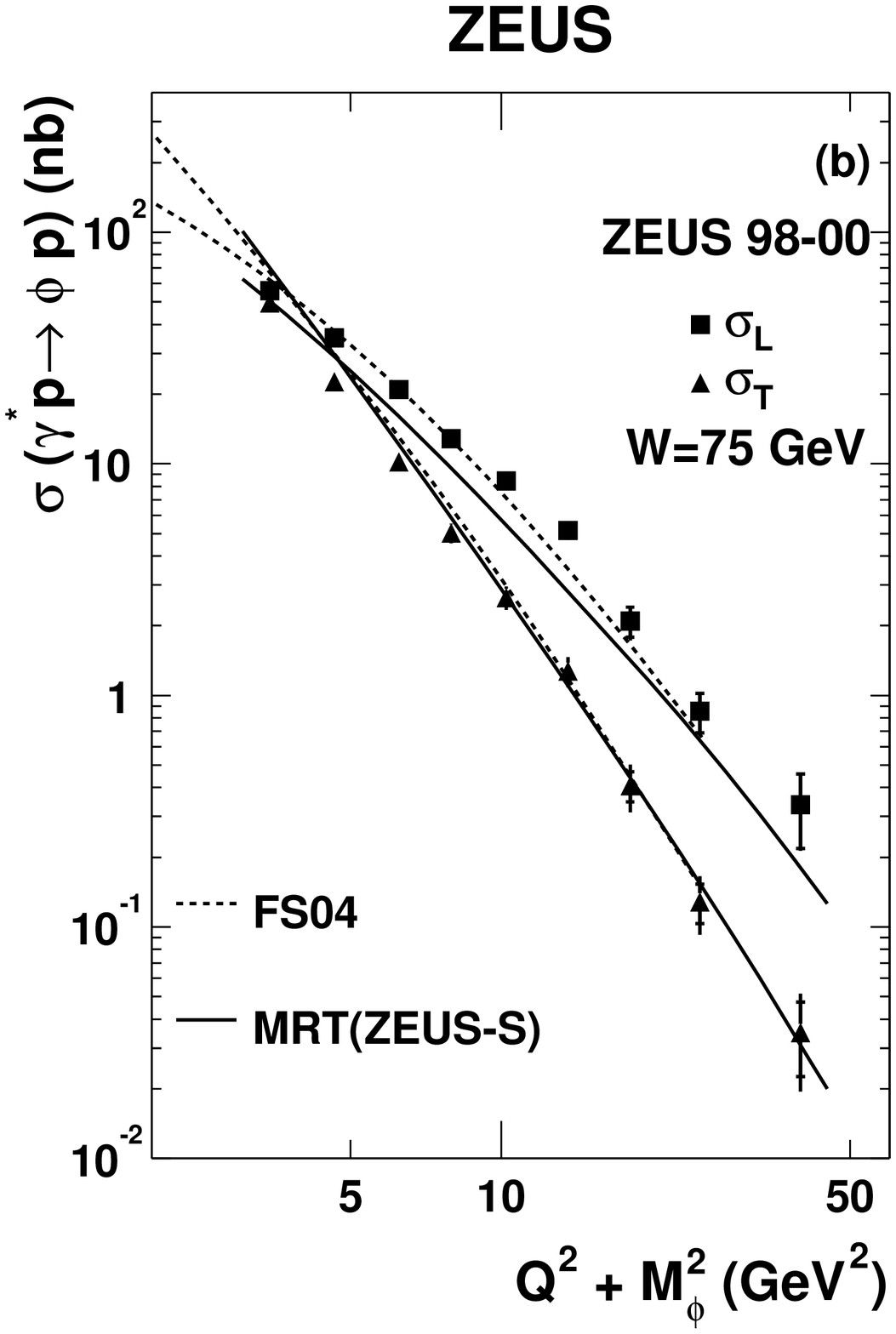,width=8.cm}
\end{center}
\caption{
(a) Exclusive $\phi$ cross section 
as a function of $W$ for three values of $Q^2$ compared with predictions from the
MRT and FS04 model.  Different gluon densities (ZEUS-S, MRST99, CTEQ6M) 
were employed in the MRT predictions as indicated in the
figure.
Note that the data at $Q^2=3.8$~GeV$^2$ are not shown 
in the figure for clarity.
(b) 
 Exclusive $\phi$ cross section 
as a function of $Q^2+M_{\phi}^2$ for $W=75 \gev$.  The data are compared to the MRT
and FS04 predictions.  The ZEUS-S gluon density was used in the MRT model.}
\label{fig-MRTcomp}
\vfill
\end{figure}

The predictions of the MRT model~\cite{pr:d62:14022} were compared to the data
using three different gluon densities as shown in 
Fig.~\ref{fig-MRTcomp}a.  
%The MRST99 and ZEUS-S gluon densities yield good agreement with the data
%at small $Q^2$ but not at higher $Q^2$, whereas the CTEQ6M gluon density
%agrees better at the higher $Q^2$ than at low $Q^2$.
The predicted $W$ dependence 
with the ZEUS-S~\cite{ZEUS-S} and CTEQ6M~\cite{CTEQ6M} 
gluon densities are compatible 
with the data, whereas the
predictions using the MRST99~\cite{MRST99} 
gluon density are too steep.  The predicted
$Q^2$ dependence is too steep for all gluon densities.  

%No conclusion on the
%preferred gluon density can be derived from the normalization given the
%large theoretical uncertainty.

The FS04 prediction~\cite{FS04} is also compared to the data
in Fig.~\ref{fig-MRTcomp}a.  
The $W$ dependence seen in the data is well reproduced, although the
normalization is somewhat low at large $Q^2$ and high at low $Q^2$. 
A similar model~\cite{pr:d60:074012} which did not employ a saturated
dipole cross section showed a somewhat steeper $W$ dependence.

The expectations from the MRT and FS04 models 
for the $Q^2$ dependence are also compared
to the data in Fig.~\ref{fig-MRTcomp}b. In this case,
the ZEUS-S gluon density was chosen for the MRT model.
This prediction agrees reasonably well with the data at higher
$Q^2$.  At small $Q^2$, the predicted transverse cross section is
too high while the predicted longitudinal cross section is too low.
The FS04 prediction reproduces the data better than the MRT prediction 
for the longitudinal
component, and is very similar to the MRT prediction for the transverse
component.

%  This result is rather surprising, as
%the longitudinal cross section is expected to be more reliably calculable
%in pQCD.

% ----------------------------------------------------------------------------
%       t dependence 
% ----------------------------------------------------------------------------
\subsection{Dependence on $t$}
\label{sec-tdep}
\begin{figure}[p]
\vfill
\begin{center}
\epsfig{file=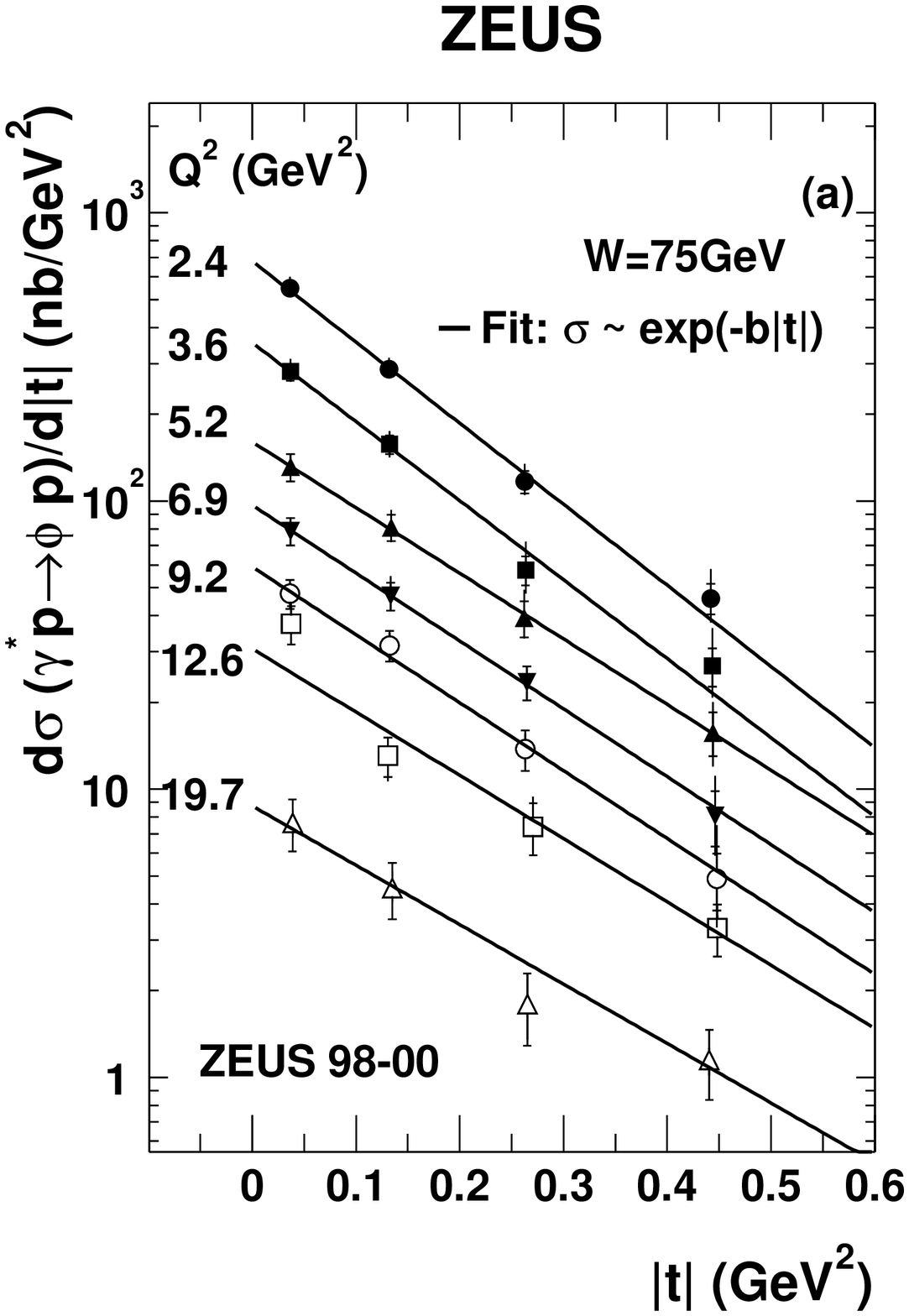,width=8.cm}\epsfig{file=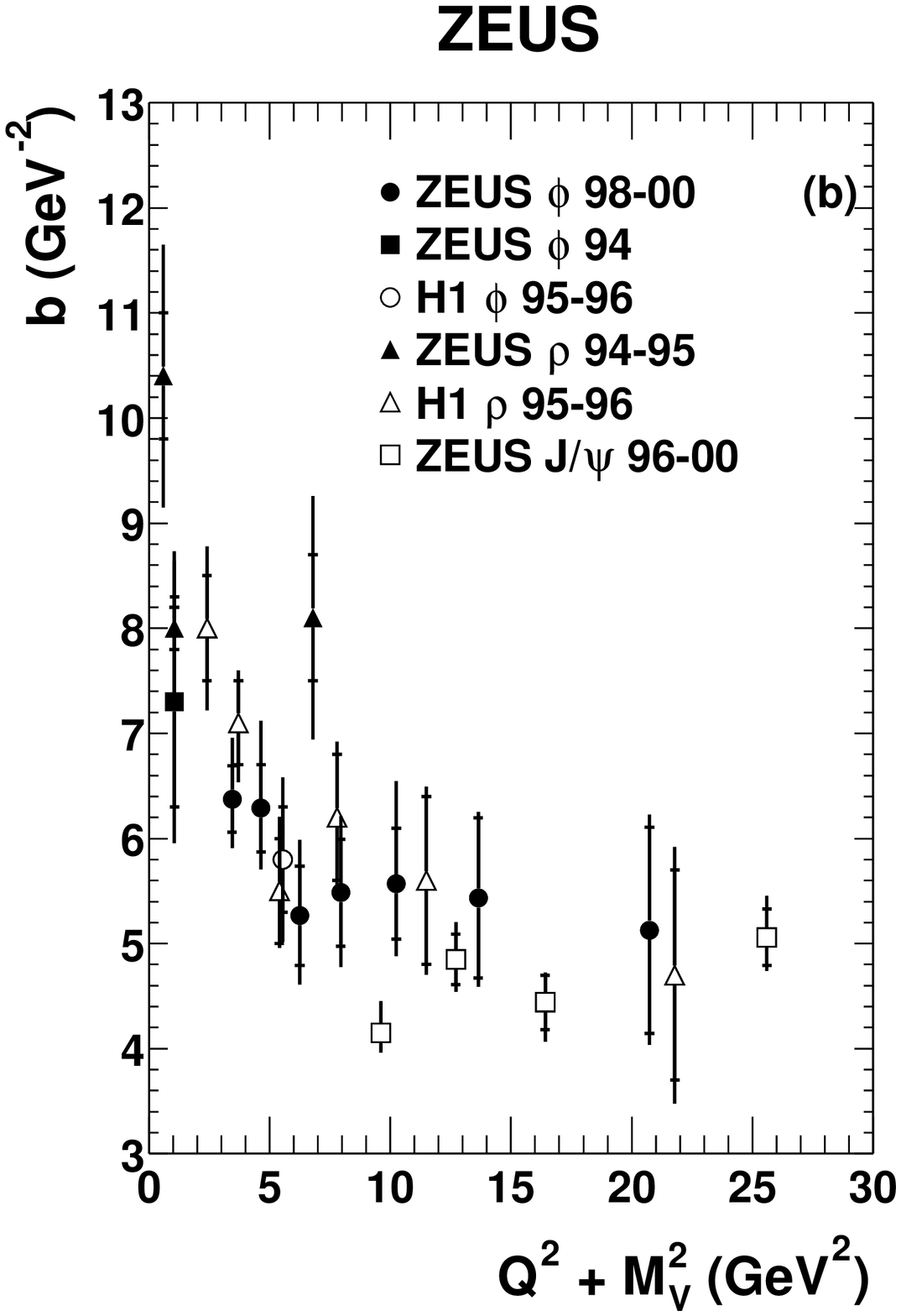,width=8.cm}
\end{center}
\caption{
(a) Differential cross-sections $d\sigma /d|t|$ 
for seven bins of $Q^2$, for $W=75 \gev$ and $|t|<0.6 \gev^2$. 
The full lines are the results of fits to the form 
$d\sigma /dt = d\sigma /dt|_{t=0} \cdot e^{bt}$.
 The overall normalization uncertainty 
of $2.8$~\% is not included in the error bar.
(b) The slope $b$, as a function of $Q^2+M_V^2$, 
compared to other ZEUS and H1 results.
The inner error bars represent the statistical uncertainty;
the outer bars are the statistical and systematic uncertainties 
added in quadrature. 
}
%\protect{~\cite{bib-elff}} 
\label{fig-dsigmadt}
\vfill
\end{figure}

The differential cross section, $d\sigma_{\rm tot}^{\gamma^* p \rightarrow \phi
p} /d|t|$, measured as a function of $t$ in the range $|t|< 0.6\gev^2$,
is shown in Fig.~\ref{fig-dsigmadt}a for different values of
$Q^2$ and $W=75 \gev$.  A function of the
form $d\sigma /dt = d\sigma /dt|_{t=0} \cdot e^{bt}$ was fitted to
the data and the results of the fit are given in
Fig.~\ref{fig-dsigmadt}b, along with measurements from other
vector mesons~\cite{
epj:c2:247, % rho - photo
epj:c6:603, % rho 96
epj:c13:371,% h1-rho 95-95
pl:b377:259, % phi- photo
pl:b483:360, % h1 - phi
epj:c24:345, % jpsiphoto
Chekanov:2004mw %DIS J/psi
}.  
The slope parameters from this analysis are given in
Table~\ref{tab:b-Q2}.
The values of $b$ from this analysis
show no $Q^2$ dependence within the present measurement
uncertainties.
The measurements in Fig.~\ref{fig-dsigmadt}b are presented as
a function of $Q^2 +M_V^2$ and are found to scale with this variable
within the present data uncertainties.

%Soft diffractive processes are described by Regge 
%phenomenology~\cite{collins:1977:regge}
%in terms of the exchange of a Pomeron trajectory.
%In hard interactions, where Regge phenomenology may not be applicable,
%an effective Pomeron trajectory may nevertheless be extracted.
The data sample was analyzed to determine the $W$ dependence as a function
of $t$. In the Regge formalism, the differential cross section can be 
expressed as
\begin{equation}
\label{regge1}
d\sigma/dt \propto W^{4(\alpha_{\pom}(t)-1)},
\end{equation}
where $\alpha_{\pom}$, the Pomeron trajectory, is usually parametrized as
\begin{equation}
\label{alpha(t)}
\alpha_{\pom}(t)= \alpha_{\pom}(0)+\alpha_{\pom}^\prime t.
\end{equation}

The trajectory measured in soft diffractive processes is
$\alpha_{\pom} =1.08 + 0.25 ~t$~\cite{pl:b348:213,pr:d10:160}.  In contrast,
$\alpha_{\pom}^\prime$ is much smaller in $J/\psi$
photoproduction: $\alpha_{\pom} =1.20 + 0.115 ~t$~\cite{epj:c24:345}.
In this notation, $\alpha_{\pom}=1 + \delta/4$.
The values of $\alpha_{\pom}(0)$ and $\alpha_{\pom}^\prime$ were determined 
by fitting the $W$ dependence of the differential cross section at different
$|t|$ values using Eq.~(\ref{regge1}) for  $Q^2=5\gev^2$. 
Since the proton-dissociative process was found to have the same $W$
dependence as the exclusive process, the extraction of $\alpha_{\pom}$
is not sensitive to this background contribution. The analysis was 
therefore extended up to $|t|=1\gev^2$.
The fit results are shown in Fig.~\ref{fig-regge}a and are
given in Table~\ref{tab:delta-t}.
The parameters of the trajectory were determined from a fit of 
Eq.(\ref{alpha(t)}) to the extracted $\alpha_{\pom}(t)$ values, as
shown in Fig.~\ref{fig-regge}b.  

The results are:
\begin{align*}
& \alpha_{\pom}(0)= 1.10\pm 0.02({\rm stat.})\pm 0.02({\rm syst.});\\
& \alpha_{\pom}^\prime= 0.08\pm0.09({\rm stat.})\pm0.08({\rm syst.})\gev^{-2}.
\end{align*}
The value of $\alpha_{\pom}^\prime$ is closer to the value
measured in $J/\psi$ production than that measured in soft diffractive
processes.

% ----------------------------------------------------------------------------
%       density matrix elements
% ----------------------------------------------------------------------------
\subsection{Decay angular distributions}
The angular distributions of the decay of the $\phi$
provide information about the photon and $\phi$ polarization states.
In the helicity frame~\cite{np:b61:381},
the production and decay of the $\phi$ 
%into a pair of oppositely charged particles 
can be described in terms of three angles: 
$\Phi_h$,  the angle between the $\phi$ production plane
and the lepton scattering plane; 
$\theta_h$, the polar angle, 
and $\phi_h$, the azimuthal angle of the positively charged kaon. 
Under the assumption of $s$-channel helicity conservation (SCHC), 
the normalized angular distribution depends only on two angles,
$\theta_h$ and $\psi_h = \phi_h - \Phi_h$.  The $\theta_h$ distribution
can be expressed in the form
\begin{equation}\label{eq-angtheta}
\frac{1}{N}\frac{dN}{d\cos \theta_h} =
\frac{3}{8}  \left[  1 + r^{04}_{00} + 
(1-3r^{04}_{00})\cos ^2 \theta_h  \right]\;\; .
\end{equation}
The spin-density matrix-element $r^{04}_{00}$ represents the probability 
that the  $\phi$  is produced in the helicity-0 state 
from a virtual photon of helicity 0 or 1. 

\begin{figure}[p]
\vfill
\begin{center}
\epsfig{file=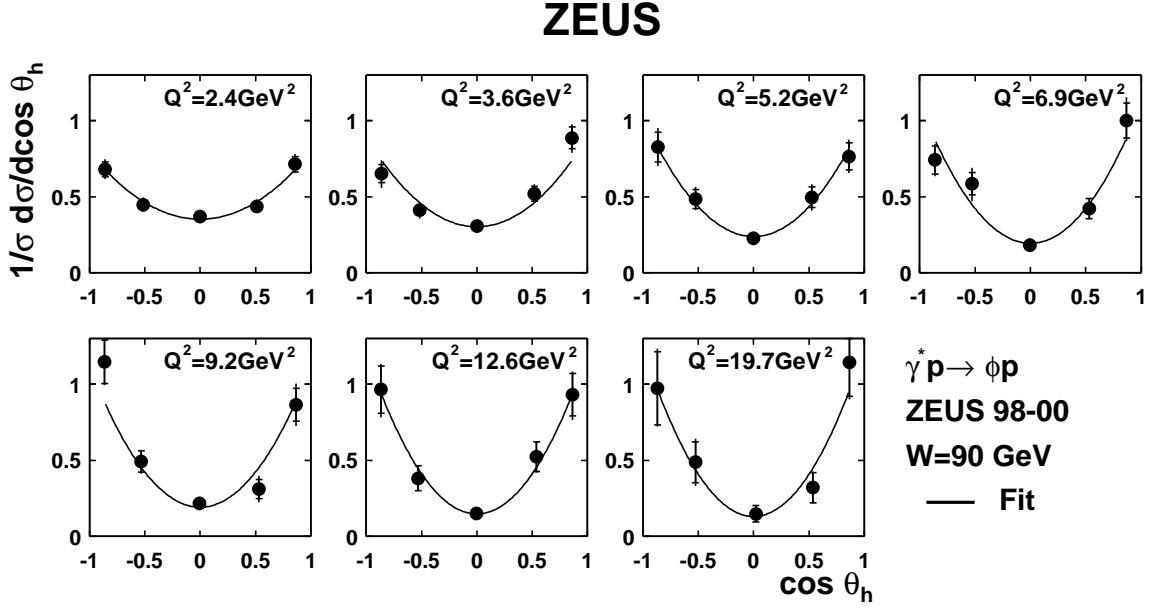,width=16.cm}
\end{center}
\caption{Normalized distributions of $\cos \theta_h$ in seven $Q^2$
 bins; the curves are the fits to Eq.~(\ref{eq-angtheta}).  The inner
 error bars represent the statistical uncertainty; the outer bars are the
 statistical and systematic uncertainties added in quadrature. 
}
\label{fig-angular}
\vfill
\end{figure}

\begin{figure}[p]
\vfill
\begin{center}
\epsfig{file=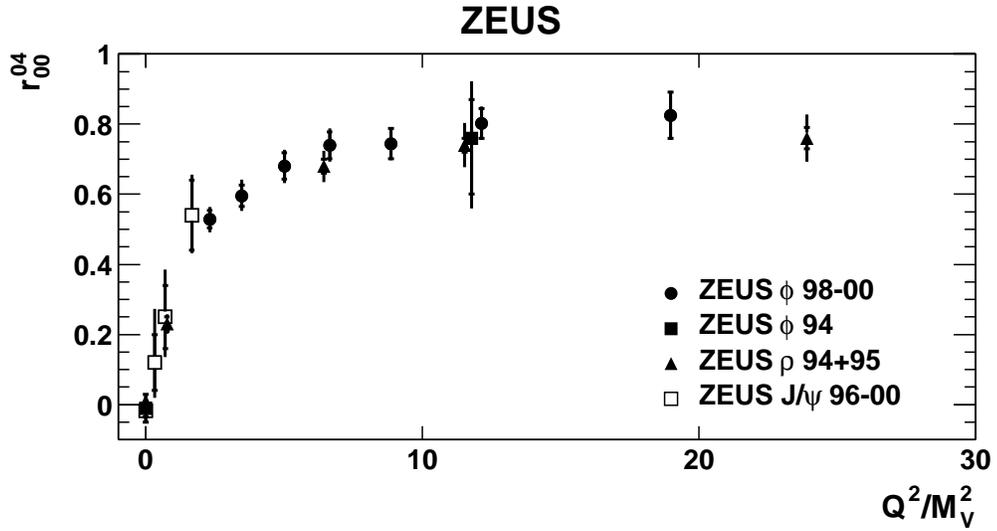,width=15.cm}
\end{center}
\caption{The values of $r^{04}_{00}$ extracted from the fits
shown in Fig.~\ref{fig-angular} are plotted as a function of $Q^2/M_V^2$
and compared with values for other vector mesons.
}
\label{fig-r04}
\vfill
\end{figure}

The normalized cross section $\frac{1}{\sigma} \frac{d\sigma_{\rm tot}^{\gamma^* p 
\rightarrow \phi p}}{d\cos(\theta_h)}$
at $W=90\gev$ is shown in Fig.~\ref{fig-angular} for seven values of
$Q^2$.  The data were fitted to Eq.~(\ref{eq-angtheta}).  
The values of $r^{04}_{00}$,
determined from the fits, are given in Table~\ref{tab:R} and 
plotted as a function
of $Q^2/M_V^2$ in Fig.~\ref{fig-r04}, where they are 
compared to the values extracted for other vector mesons~\cite{
epj:c2:247, % rho - photo
epj:c6:603, % rho 96
epj:c12:393, % zeus - helicity rho
pl:b377:259, % phi- photo
pl:b380:220, % phi94
epj:c24:345, % jpsiphoto
Chekanov:2004mw %DIS J/psi
}.  The values
of $r^{04}_{00}$ for the different vector mesons are found to scale 
in this variable.

The values of $r^{04}_{00}$  were also extracted as a function of $W$
for two different $Q^2$ values, and these are given
in Table~\ref{tab:RW}.  No dependence on $W$ was observed.

\subsubsection{Longitudinal and transverse cross sections}
\label{section-R}
The ratio of the longitudinal to transverse cross section,
$R=\sigma_L/\sigma_T$, was calculated as a function of $Q^2$
from $r^{04}_{00}$ according to the relation
\begin{equation}
\label{eq:Rfromr04}
R = \frac{1}{\epsilon}  \frac{r^{04}_{00}}{1-r^{04}_{00}},
\end{equation}
which is valid under the assumption of SCHC\footnote{The validity of SCHC
has been tested for other vector mesons 
and found to be adequate for the purpose of
extracting $R$~\cite{epj:c12:393}.}.  The average value of 
$\epsilon = 0.99$ was used in extracting $R$.  The
values of $R$ are presented as a function of $Q^2$ 
in Table~\ref{tab:R} and as a function of $W$ in Table~\ref{tab:RW}.
The scaling of $r^{04}_{00}$ with  $Q^2/M_V^2$ implies that $R$ also
scales in this variable.  This is expected in dipole models
of diffraction~\cite{barone:predazzi}.

\begin{figure}[p]
\vfill
\begin{center}
\epsfig{file=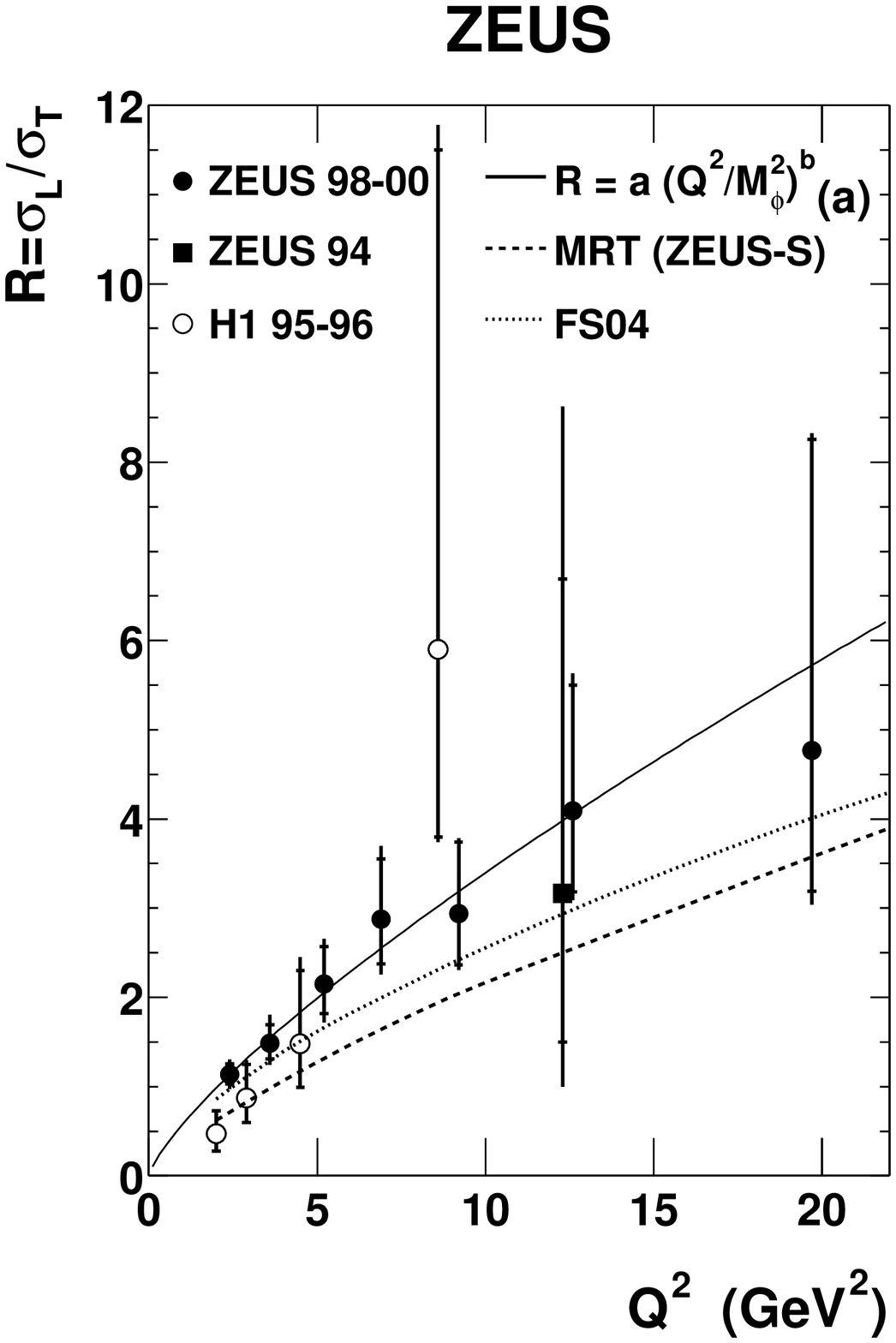,width=8.cm}\epsfig{file=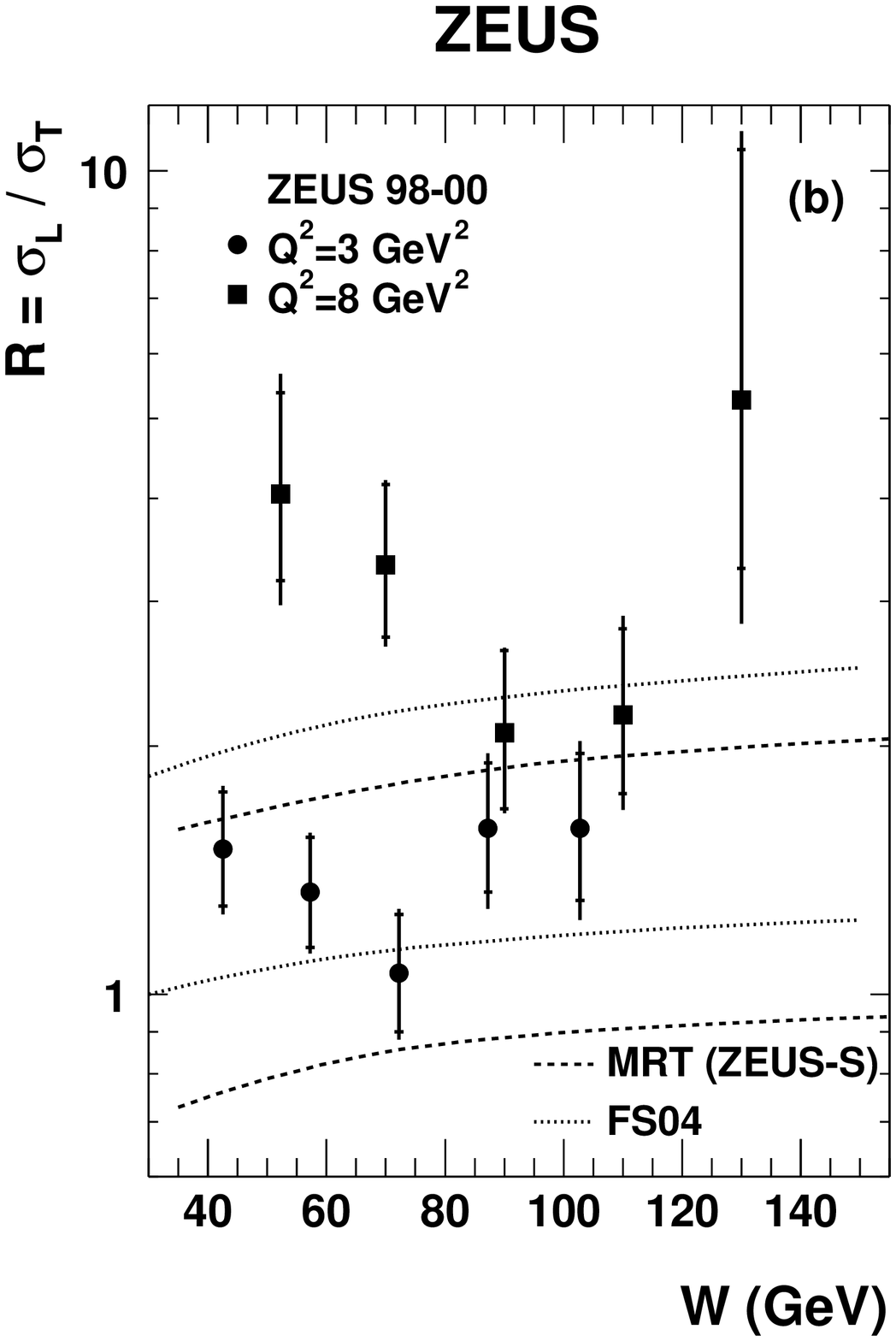,width=8.cm}
\end{center}
\caption{ (a) Ratio $R=\sigma_L/\sigma_T$ as a function of $Q^2$ for
exclusive $\phi$ production;
the full line is the result of the two-parameter fit as shown in the
plot.
(b) Ratio $R=\sigma_L/\sigma_T$ as a function of $W$ extracted in this 
analysis for two different $Q^2$ values.
The dashed curve in both plots 
is from the MRT model with the ZEUS-S gluon density,
while the dotted curve is from the FS04 model. 
The inner error bars show the statistical uncertainty, the outer the
statistical and systematic uncertainties added in quadrature.
}
\label{fig-RvsW}
\vfill
\end{figure}

The values of $R$  are compared with previous
ZEUS~\cite{pl:b377:259,pl:b380:220} and H1~\cite{pl:b483:360} 
results in Fig.~\ref{fig-RvsW}a
as a function of $Q^2$ and in Fig.~\ref{fig-RvsW}b as a function of $W$. The
$Q^2$ dependence is
well described by the expression $R=a(Q^2/M_{\phi}^2)^b$. The
parameters were extracted from the fit to $r^{04}_{00}$ data for which
the statistical errors are Gaussian, yielding $a=0.51\pm0.07({\rm
stat.})\pm0.05({\rm syst.})$ and $b=0.86\pm0.11({\rm
stat.})\pm0.05({\rm syst.})$.  The prediction from the MRT model (using the
ZEUS-S gluon density) is also
shown in Fig.~\ref{fig-RvsW}, as is the FS04 prediction.  
The general power-law dependence of
$R$ with $Q^2$ is reproduced, but the model predictions 
systematically underestimate
the measurements. The dipole-model prediction (FS04) is in somewhat better
agreement with the ZEUS data than the MRT-model prediction
(with the ZEUS-S gluon density), particularly at the lower $Q^2$ values.
The weak dependence of $R$ on $W$ observed in Fig.~\ref{fig-RvsW}b is
  consistent with both the MRT and FS04 models.
% ----------------------------------------------------------------------------
%       Conclusions
% ----------------------------------------------------------------------------
\section{Summary}\label{sec-conclusions}
The exclusive electroproduction of $\phi$ mesons, 
$e\: p\rightarrow e \:\phi\:p$, 
has been measured with the ZEUS detector at HERA 
for photon virtualities in the range
$2<Q^2<70\gev^2$, 
for photon-proton center-of-mass energies in the range $35<W<145\gev$ 
and for four-momentum-transfer squared in the range $|t|<0.6\gev^2$. 
The extracted $\gamma^*\:p$ cross section 
rises with $W$ as $\sigma \propto W^{\delta}$, 
with a slope parameter $\delta \approx 0.4$. This value is between
the `soft' diffraction value and the value observed in $J/\psi$ production. No
$Q^2$ or $t$ dependence of $\delta$ was observed within the present 
experimental precision of the data.

The high-precision data from this analysis 
reveal that the $Q^2$ dependence of the cross section
cannot be fitted with a single power over the measured $Q^2$ range.
The longitudinal and transverse components of the cross section were
separately extracted using the measured value of $R$ 
and the different $Q^2$ dependence of the cross section for the two
helicity components is clearly seen.

The $t$ distribution, measured for $|t|<0.6\gev^2$, is well described by an 
exponential dependence over the range $2<Q^2<70\gev^2$. 
The slope ranges from $6.4 \pm 0.4 \gev^{-2}$ at 
$Q^2=2.4 \gev^2$ to $5.1 \pm 1.1 \gev^{-2}$ at $Q^2=19.7 \gev^2$.

%, and
%is consistent with that observed for other vector mesons when presented as
%a function of $Q^2+M_V^2$.

The ratio of the cross sections for longitudinally and transversely
polarized photons, $R$, increases with $Q^2$ and can be fitted
by a power-law dependence.

The data from this analysis were compared to previously published 
$e\: p \rightarrow e\: V\: p$ data.  
The values of $\delta$ and $b$ were found to scale, within the 
present accuracy, when plotted as
a function of $Q^2 + M_V^2$.  The ratio of longitudinal to transverse
cross sections was seen to scale with $Q^2/M_V^2$.

The MRT model does not
reproduce the $Q^2$ dependence of the $\gamma^*p$ cross section
observed in the data, while the FS04 prediction reasonably reproduces 
the data for the longitudinal photon polarization.
This conclusion is also reflected in the better agreement
of FS04 with the data for the measurement of $R$ vs $Q^2$.  The
$W$ dependence of the data can be reproduced in both models.

\section*{Acknowledgements}
We thank the DESY directorate for their strong support and encouragement.
The special effort of the HERA group is gratefully acknowledged.
The design, construction and installation of the ZEUS detector 
has been made possible by the efforts of many people 
who are not listed as authors.  Special thanks go to Thomas Teubner for
providing us with the predictions from the MRT model, and to Jeff
Forshaw for providing us with the predictions from the FS model.

\vfill\eject

%------------------------------------------------------------------------------
%       Bibliography
%------------------------------------------------------------------------------
\providecommand{\etal}{et al.\xspace}
\providecommand{\coll}{Coll.\xspace}
\catcode`\@=11
\def\@bibitem#1{%
\ifmc@bstsupport
  \mc@iftail{#1}%
    {;\newline\ignorespaces}%
    {\ifmc@first\else.\fi\orig@bibitem{#1}}
  \mc@firstfalse
\else
  \mc@iftail{#1}%
    {\ignorespaces}%
    {\orig@bibitem{#1}}%
\fi}%
\catcode`\@=12

%------------------------------------------------------------------------------
%       Tables
%------------------------------------------------------------------------------

%-------------------------------------------------------------------------------
%       Tables
%-------------------------------------------------------------------------------
\newcommand{\normerror}{{$^{+4.1}_{-5.3}$~\%} }

\begin{table}[ht]
\begin{center}
\begin{tabular}{|c|c|c|c|c|c|c|}
\hline
$Q^2$ range & $Q^2$& $W$ range & $W$ & $Events$ & $A$ & $\sigma^{\gamma^{*}p\rightarrow\phi p}_{\rm tot} $  \\
($\gev^2$) & ($\gev^2$) & ($\gev$) & ($\gev$) &  & $(\%)$ &  (nb)  \\
\hline
 &  & 35--45 & 40  & 203  & 19.4  & $76.4\pm 6.5 ^{+4.0}  _{-5.9} $\\
 &  & 45--55 & 50  & 255  & 18.7  & $101.2\pm 7.7 ^{+8.1}  _{-4.2} $\\
 &  & 55--65 & 60  & 220  & 19.6  & $101.9\pm 8.3 ^{+6.3}  _{-7.4} $\\
 2--3 & 2.4  & 65--75 & 70  & 210  & 19.6  & $112.8\pm 9.4 ^{+6.3}  _{-7.6} $\\
 &  & 75--85 & 80  & 167  & 19.3  & $107\pm 11 ^{+4}  _{-6} $\\
 &  & 85--95 & 90  & 175  & 20.7  & $122\pm 11 ^{+10}  _{-10} $\\
 &  & 95--105 & 100  & 134  & 19.5  & $110\pm 11 ^{+8}  _{-7} $\\
\hline
 &  & 40--50 & 45  & 184  & 20.9  & $47.0\pm 4.1 ^{+1.4}  _{-2.5} $\\
 &  & 50--60 & 55  & 146  & 21.3  & $44.3\pm 4.5 ^{+2.8}  _{-3.6} $\\
 3--5 & 3.8 & 60--70 & 65  & 169  & 22.3 & $56.7\pm 5.1 ^{+3.6}  _{-4.0} $\\
 &  & 70--85 & 77.5  & 218  & 22.0  & $62.3\pm 5.0 ^{+4.3}  _{-4.2} $\\
 &  & 85--100 & 92.5  & 158  & 21.3  & $57.4\pm 5.4 ^{+3.2}  _{-4.3} $\\
 &  & 100--115 & 107.5  & 123  & 21.2  & $59.0\pm 6.2 ^{+3.0}  _{-2.6} $\\
\hline
 &  & 45--55 & 50  & 111  & 34.6  & $16.4\pm 1.8 ^{+1.1}  _{-1.1} $\\
 &  & 55--70 & 62.5  & 168  & 37.6  & $19.1\pm 1.7 ^{+1.5}  _{-1.4} $\\
 5--9 & 6.5 & 70--85 & 77.5  & 136  & 38.5  & $19.6\pm 1.9 ^{+1.6}  _{-1.5} $\\
 &  & 85--100 & 92.5  & 123  & 37.8  & $21.6\pm 2.3 ^{+1.1}  _{-1.1} $\\
 &  & 100--115 & 107.5  & 116 & 40.9  & $23.1\pm 2.5 ^{+1.3}  _{-1.3} $\\
 &  & 115--135 & 125  & 70  & 44.6  & $25.3\pm 3.5 ^{+3.9}  _{-3.9} $\\
\hline
 &  & 50--60 & 55  & 64.5  & 58.6  & $5.05\pm 0.73 ^{+0.37}  _{-0.37} $\\
 &  & 60--75 & 67.5  & 88.6  & 57.0  & $4.96\pm 0.59 ^{+0.27}  _{-0.26} $\\
 9--30 & 13.0 & 75--90 & 82.5  & 87.3  & 56.5  & $6.12\pm 0.81 ^{+0.34}  _{-0.40} $\\
 &  & 90--105 & 97.5  & 86.6  & 55.6  & $6.82\pm 0.85 ^{+0.40}  _{-0.37} $\\
 &  & 105--125 & 115  & 85.8  & 58.0  & $6.33\pm 0.78 ^{+0.50}  _{-0.22} $\\
 &  & 125--145 & 135  & 61.7  & 56.8  & $6.65\pm 0.97 ^{+0.44}  _{-0.52} $\\
\hline

\end{tabular}
\caption{The $\gamma^{*}p\rightarrow\phi p$ cross section,
$\sigma^{\gamma^{*}p\rightarrow\phi p}_{\rm tot}$, as a function of
  $W$ and $Q^2$.  The first uncertainty is statistical and the second
  systematic. The overall normalization error of \normerror is not
  included. The $Q^2$ and $W$ values at which the 
  cross section is evaluated are given in the second and fourth columns. 
  The number of extracted events corrected for
  non-resonant background and the acceptance, A, are given for each
  measurement.}
\label{tab:xsecW}
\end{center}
\end{table}

\begin{table}[ht]
\begin{center}
\begin{tabular}{|c|c|c|}
\hline
 $Q^2$ range & $Q^2$ & $\delta$  \\
 ($\gev^2$) & ($\gev^2$) & $(\sigma\propto W^\delta)$  \\
\hline
 2--3 & 2.4  & $0.41\pm 0.10 ^{+0.06}  _{-0.05} $\\
 3--5 & 3.8  & $0.34\pm 0.13 ^{+0.05}  _{-0.03} $\\
 5--9 & 6.5  & $0.43\pm 0.15 ^{+0.13}  _{-0.13} $\\
 9--30 & 13.0  & $0.38\pm 0.18^{+0.07}  _{-0.06} $\\
\hline
\end{tabular}
\caption{The $\delta$ parameter ($\sigma\propto W^\delta$) as a function of
$Q^{2}$. The first uncertainty is statistical and the second
systematic.}
\label{tab:delta-q2}
\end{center}
\end{table}

\begin{table}[ht]
\begin{center}
\begin{tabular}{|c|c|c|c|c|}
\hline
 $Q^2$ range & $Q^2$ & $Events$ & $A$ & $\sigma^{\gamma^{*}p\rightarrow\phi p}_{\rm tot} $  \\
 ($\gev^2$) & ($\gev^2$) &  & $(\%)$ &  (nb)  \\
\hline
2--3 & 2.4  & 1389  & 19.3  & $105.5\pm 3.4 ^{+4.6}  _{-6.0} $\\
3--4.5 & 3.6  & 842  & 21.4  & $57.6\pm 2.4 ^{+3.2}  _{-3.5} $\\
4.5--6 & 5.2  & 420  & 26.6  & $31.1\pm 1.8 ^{+1.7}  _{-1.8} $\\
6--8 & 6.9  & 376  & 42.5  & $17.9\pm 1.1 ^{+1.0}  _{-1.0} $\\
8--11 & 9.2  & 314  & 52.1  & $11.06\pm 0.73 ^{+0.56}  _{-0.51} $\\
11--15 & 12.6  & 200  & 59.9  & $6.42\pm 0.52 ^{+0.24}  _{-0.18} $\\
15--20 & 17.1  & 61.8  & 55.2  & $2.50\pm 0.37 ^{+0.16}  _{-0.22} $\\
20--30 & 24.0  & 32.3  & 62.5  & $0.98\pm 0.19 ^{+0.05}  _{-0.05} $\\
30--70 & 38.8  & 10.5  & 53.5  & $0.37\pm 0.13 ^{+0.04}  _{-0.04} $\\
\hline
\end{tabular}
\caption{The $\gamma^{*}p\rightarrow\phi p$ cross section,
$\sigma^{\gamma^{*}p\rightarrow\phi p}_{\rm tot}$, as a function of 
  $Q^2$ for $W=75\gev$. The second column gives the value of $Q^2$ at which the cross section is
quoted. The number of
  extracted events corrected for non-resonant background and the
  acceptance, A, are given for each measurement. The first uncertainty
  is statistical and the second systematic. The overall normalization
  error of \normerror is not included.}
\label{tab:xsecQ2}
\end{center}
\end{table}

\begin{table}[ht]
\begin{center}
\begin{tabular}{|c|c|c|c|}
\hline
  $Q^2$ range & $Q^2$ & $b \, (\frac{d\sigma}{d|t|}\propto e^{-b|t|})$ & $\left.\frac{d\sigma}{d|t|}\right|_{t=0}$  \\
($\gev^2$) & ($\gev^2$) & ($\gev^{-2}$) & (nb/$\gev^2$) \\
\hline

2--3 & 2.4  & $6.37 \pm 0.32 ~^{+0.27}  _{-0.28} $ &$678 \pm 37 ~^{+39}  _{-56} $\\
3--4.5 & 3.6  & $6.29 \pm 0.42 ~^{+0.45}  _{-0.32} $ & $352 \pm 26 ~^{+27}  _{-22} $ \\
4.5--6 & 5.2  & $5.26 \pm 0.48 ~^{+0.28}  _{-0.37} $ & $162 \pm 16 ~^{+9}  _{-13} $\\
6--8 & 6.9  & $5.49 \pm 0.52 ~^{+0.33}  _{-0.47} $ & $98 \pm 10 ~^{+8}  _{-10} $\\
8--11 & 9.2  & $5.58 \pm 0.54 ~^{+0.45}  _{-0.32} $ & $61.3 \pm 6.6 ~^{+4.8}  _{-3.7} $\\
11--15 & 12.6  & $5.45 \pm 0.78 ~^{+0.21}  _{-0.37} $ & $32.9 \pm 5.5 ~^{+1.6}  _{-3.2} $\\
15--30 & 19.7  & $5.10 \pm 0.98 ~^{+0.43}  _{-0.40} $ & $8.8 \pm 1.9 ~^{+0.8}  _{-1.0} $\\
\hline

\end{tabular}
\caption{The slope parameter, $b$,
as a function of $Q^{2}$ for $W=75\gev$. The cross section is quoted at
the $Q^2$ value given in the second column.  The first uncertainty is
statistical and the second systematic.}
\label{tab:b-Q2}
\end{center}
\end{table}

\begin{table}[ht]
\begin{center}
\begin{tabular}{|c|c|c|c|}
\hline
$|t|$ range & $|t|$ & $\delta$ &  $\alpha_{I\!P}$\\
($\gev^2$) & ($\gev^2$) & $(\sigma\propto W^\delta)$ & $\alpha_{I\!P} = 1+\delta/4$ \\
\hline

 0--0.08& 0.025  & $0.34\pm 0.11 ^{+0.08}  _{-0.08} $ & $1.085 \pm 0.027 ^{+0.021}  _{-0.020} $\\
 0.08--0.20& 0.12  & $0.39\pm 0.12 ^{+0.09}  _{-0.08} $ & $1.098 \pm 0.030 ^{+0.022} _{-0.021} $\\
 0.20--0.35& 0.25  & $0.44\pm 0.15 ^{+0.11}  _{-0.11} $ & $1.110 \pm 0.038 ^{+0.027} _{-0.028} $\\
 0.35--0.6& 0.45  & $0.31\pm 0.18 ^{+0.26}  _{-0.34} $ & $1.078 \pm 0.045 ^{+0.064} _{-0.086} $\\
 0.6--1.0& 0.73  & $-0.11\pm 0.33 ^{+0.24}  _{-0.32} $ & $0.972 \pm 0.083 ^{+0.060} _{-0.080} $\\

\hline
\end{tabular}
\caption{The $\delta$ parameter and the Pomeron
trajectory $\alpha_{I\!P}$ as a function of $|t|$. The first
uncertainty is statistical and the second systematic.}
\label{tab:delta-t}
\end{center}
\end{table}

\begin{table}[ht]
\begin{center}
\begin{tabular}{|c|c|c|c|}
\hline
$Q^2$ range & $Q^2$ & $r^{04}_{00}$ & $R=\sigma_L/\sigma_T$  \\
($\gev^2$) & ($\gev^2$) &  &\\
\hline

2--3 & 2.4  & $0.529 \pm 0.025 ^{+0.025}  _{-0.029} $  & $1.13^{+0.12}  _{-0.11} ~^{+0.12}  _{-0.12} $\\
3--4.5 & 3.6  & $0.595 \pm 0.031 ^{+0.036}  _{-0.030} $  & $1.49^{+0.20}  _{-0.18} ~^{+0.24}  _{-0.17} $ \\
4.5--6 & 5.2  & $0.680 \pm 0.038 ^{+0.026}  _{-0.031} $ & $2.15^{+0.42}  _{-0.33} ~^{+0.28}  _{-0.28} $\\
6--8 & 6.9  & $0.740 \pm 0.038 ^{+0.028}  _{-0.027} $ & $2.88^{+0.67}  _{-0.50} ~^{+0.47}  _{-0.37} $\\
8--11 & 9.2  & $0.744 \pm 0.043 ^{+0.017}  _{-0.019} $ & $2.94^{+0.80}  _{-0.57} ~^{+0.27}  _{-0.27} $\\
11--15 & 12.6  & $0.802 \pm 0.043 ^{+0.022}  _{-0.018} $& $4.09^{+1.41}  _{-0.91} ~^{+0.63}  _{-0.42} $\\
15--30 & 19.7  & $0.825 \pm 0.066 ^{+0.019}  _{-0.024} $ & $4.77^{+3.49}  _{-1.58} ~^{+0.72}  _{-0.71} $\\

\hline
\end{tabular}
\caption{The spin-density matrix-element $r^{04}_{00}$ and the ratio
of cross sections for longitudinally and transversly polarized
photons, $R$, as a function of $Q^{2}$. The first uncertainty is
statistical and the second systematic. Due to the transformation
given in Eq.~(\ref{eq:Rfromr04}), the error on the measurement of $R$ is
asymmetric.}
\label{tab:R}
\end{center}
\end{table}

\begin{table}[ht]
\begin{center}
\begin{tabular}{|c|c|c|c|c|c|}
\hline
 $Q^2$ range & $Q^2$  & $W$ range & $W$ & $ r^{04}_{00} $ & $R=\sigma_L/\sigma_T$  \\
 ($\gev^2$) & ($\gev^2$) & ($\gev$) & ($\gev$) & &\\
\hline
 &  & 35--50 & 42.5  & $0.597 \pm 0.038 ^{+0.019}_{-0.019} $ & $1.50^{+0.26}  _{-0.22} ~^{+0.12}  _{-0.12} $ \\
 &  & 50--65 & 57.25  & $0.569 \pm 0.038 ^{+0.017}_{-0.017} $ & $1.33^{+0.22}  _{-0.19} ~^{+0.10}  _{-0.09} $\\
 2--5& 3 & 65--80 & 72.25  & $0.511 \pm 0.041 ^{+0.017}_{-0.020} $ & $1.06^{+0.19}  _{-0.16} ~^{+0.08}  _{-0.08} $\\
 &  & 80--95 & 87.25  & $0.611 \pm 0.043  ^{+0.026}_{-0.028} $ & $1.59^{+0.32}  _{-0.26} ~^{+0.19}  _{-0.18} $\\
 &  & 95--115 & 102.75  & $0.612 \pm 0.048 ^{+0.033}_{-0.033} $ & $1.59^{+0.37}  _{-0.29} ~^{+0.24}  _{-0.21} $ \\
\hline
 &  & 45--60 & 52.25  & $0.800 \pm 0.042  ^{+0.031}  _{-0.029} $ & $4.05^{+1.33}  _{-0.87} ~^{+0.93}  _{-0.64} $\\
 &  & 60--80 & 70.0  & $0.767 \pm 0.038  ^{+0.015}  _{-0.017} $ & $3.32^{+0.84}  _{-0.61} ~^{+0.31}  _{-0.30} $\\
 5--20 & 8 & 80--100 & 90.0  & $0.673 \pm 0.048  ^{+0.015}  _{-0.014} $ & $2.08^{+0.54}  _{-0.40} ~^{+0.14}  _{-0.13} $\\
 &  & 100--120 & 110.0  & $0.684 \pm 0.050 ^{+0.032}  _{-0.030} $ & $2.19^{+0.59}  _{-0.43} ~^{+0.36}  _{-0.27} $\\
 &  & 120--145 & 130.0  & $0.839 \pm 0.074 ^{+0.046}  _{-0.048} $ & $5.3^{+5.3}  _{-2.0} ~^{+2.5}  _{-1.4} $\\
\hline

\end{tabular}
\caption{The spin-density matrix-element $r^{04}_{00}$ and the ratio
of cross sections for longitudinally and transversly polarized
photons, $R$, as a function of $W$ for two bins in $Q^2$. The first
uncertainty is statistical and the second systematic. Due to
the transformation given in Eq.~(\ref{eq:Rfromr04}) the error on the
measurement of $R$ is asymmetric.}
\label{tab:RW}
\end{center}
\end{table}

\clearpage
\newpage

%
%------------------------------------------------------------------------------
%       Figures
%------------------------------------------------------------------------------

\begin{figure}[p]
\vfill
\begin{center}
\epsfig{file=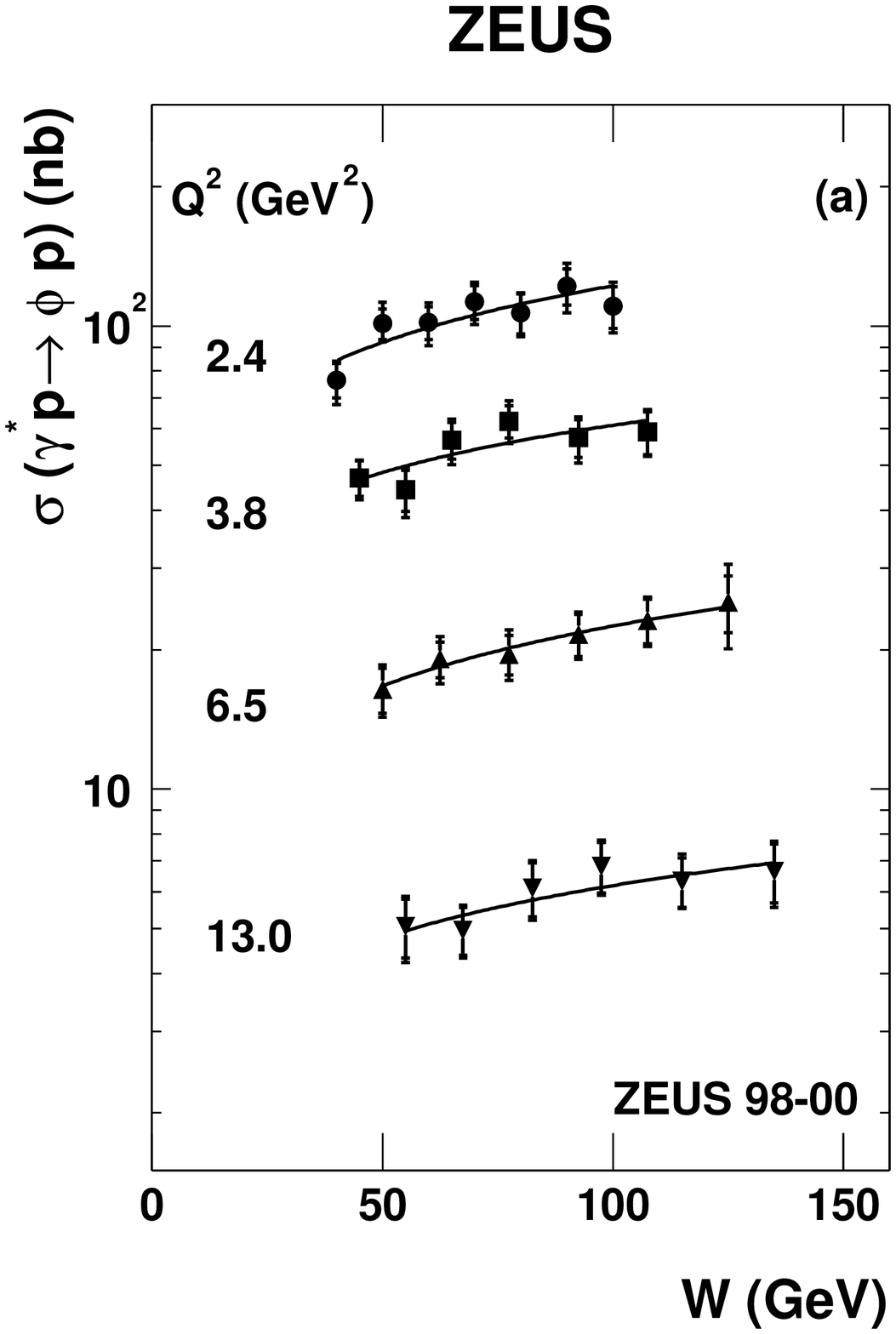,width=8.cm}\epsfig{file=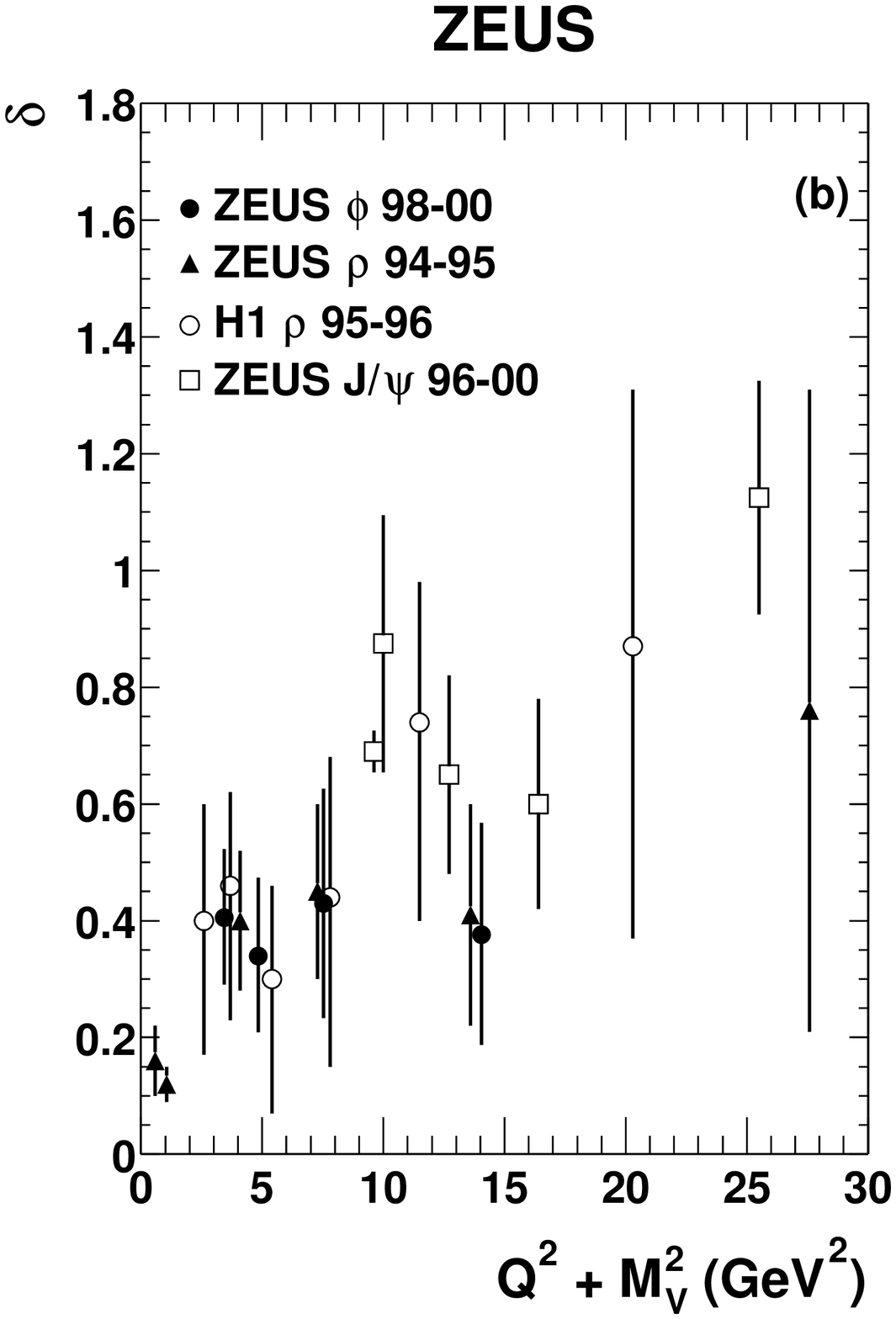,width=8.cm}
\end{center}
\caption{
(a) Exclusive $\phi$ cross section 
as a function of $W$ for four values of $Q^2$.  The solid lines are the
results of a fit to the form $\sigma \propto W^{\delta}$.
 The overall normalization uncertainty 
of \normerror is not included in the error bar.
(b) The extracted values of $\delta$ compared with results from other
vector mesons.  The error bars represent the quadratic sum of the 
statistical and systematic uncertainties.  Note that the ZEUS $\rho$ data point
at $Q^2+M_V^2$ near $8\gev^2$ has been shifted down by $0.3\gev^2$ for
clarity of presentation.}
\label{fig-xsec}
\vfill
\end{figure}

\begin{figure}[p]
\vfill
\begin{center}
\epsfig{file=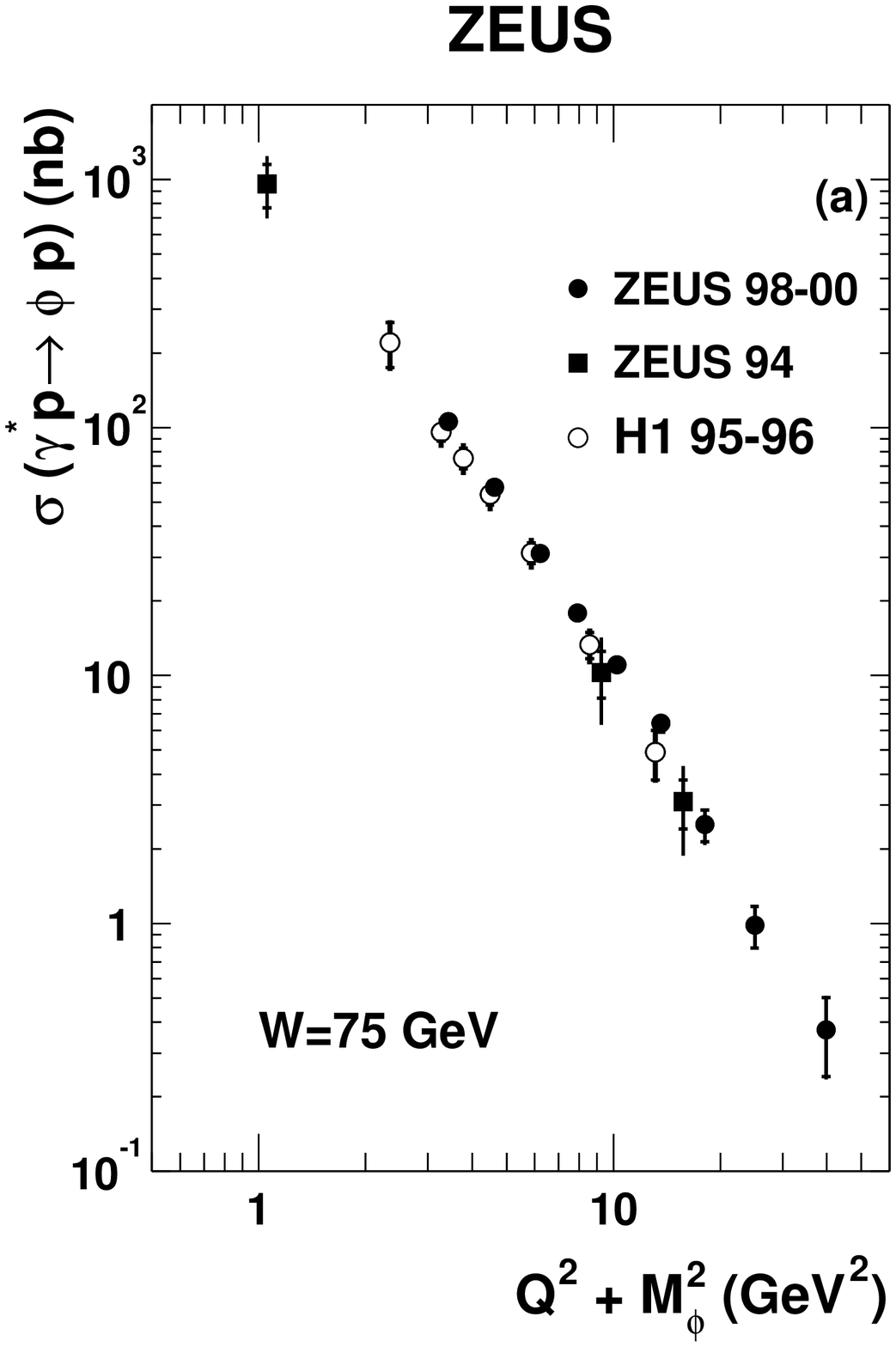,width=8cm}\epsfig{file=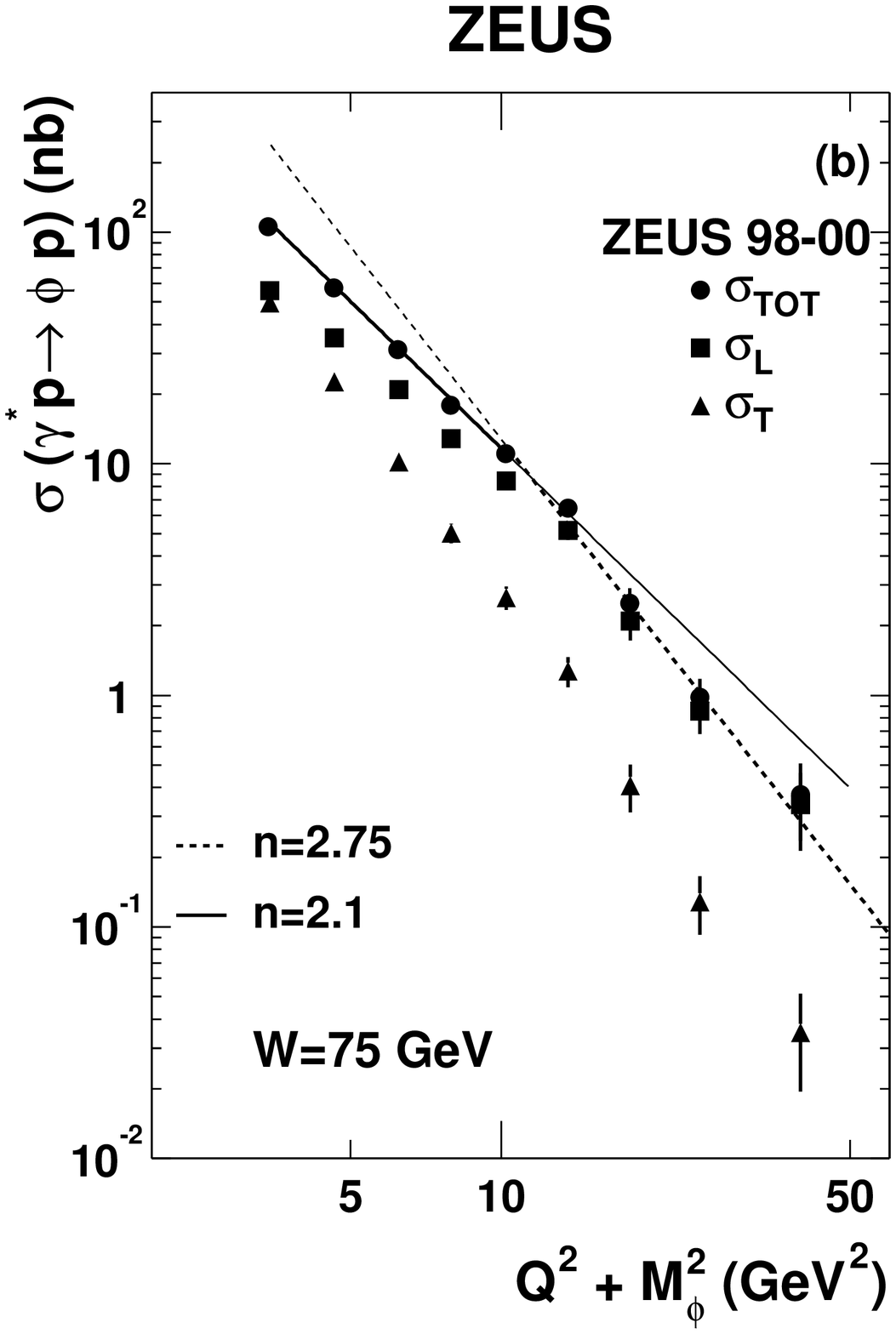,width=8.cm}
\end{center}
\caption{
 Exclusive $\phi$ cross section 
as a function of $Q^2+M_{\phi}^2$ for $W=75 \gev$.  (a) The total
cross section compared with previous measurements. 
(b) The separate longitudinal and transverse contributions to the
cross section. The curves represent the results of the fits to the
total cross section as described in the text.  
The error bars represent the quadratic sum of the statistical and
systematic uncertainties. 
The overall normalization uncertainty of \normerror is not shown.
}
\label{fig-Q2xsec}
\vfill
\end{figure}

\begin{figure}[p]
\vspace{-1cm}

\begin{center}
\epsfig{file=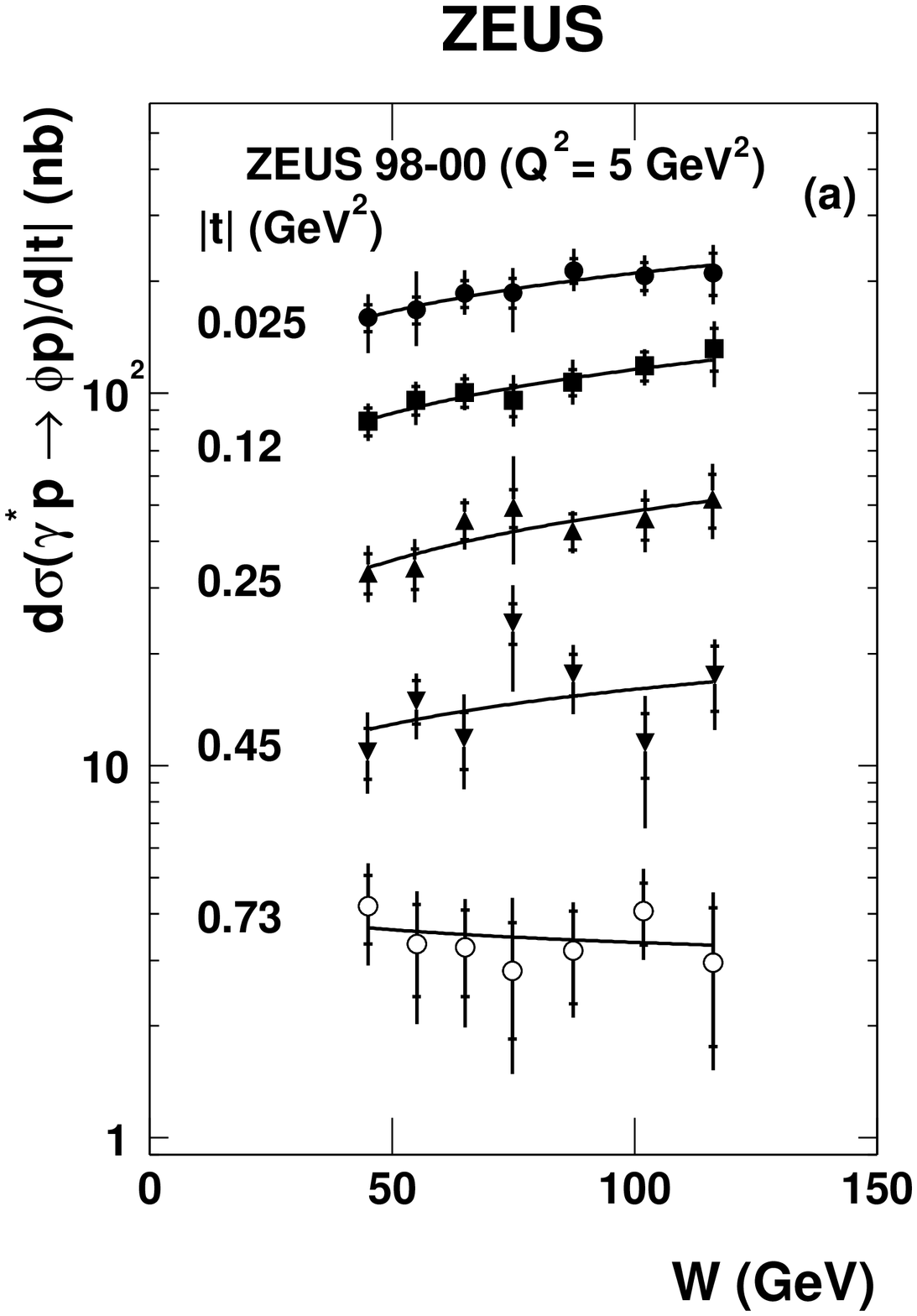,width=8.cm}\epsfig{file=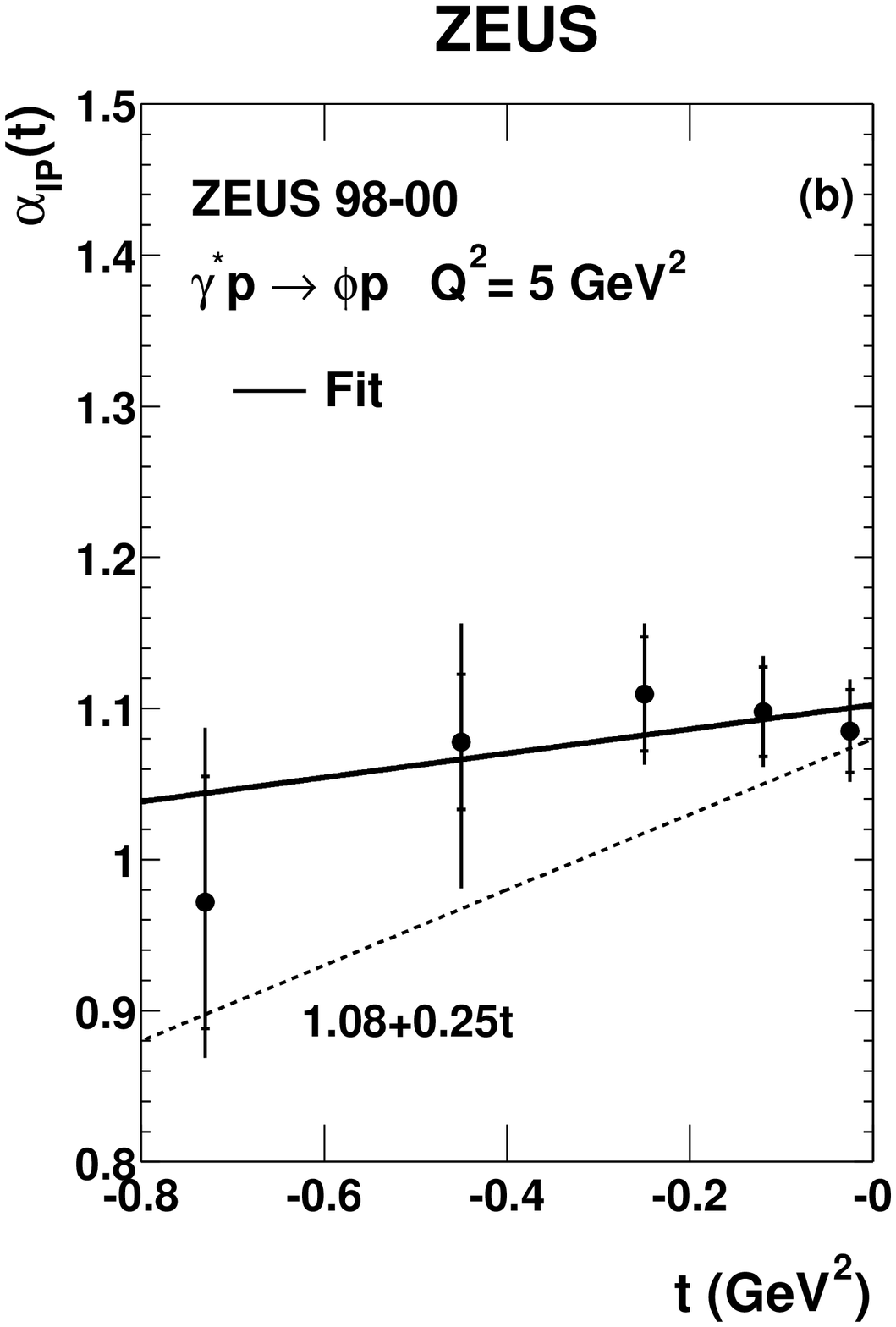,width=8.cm}
\end{center}
\caption{ (a) Differential cross-sections $d\sigma /d|t|$ as a
function of $W$ for different values of $t$ at $Q^2=5 \gev^2$. The
full lines are fits to the form $W^{\delta}$. The mean contribution from proton
dissociation has been subtracted.
The overall normalization uncertainty of \normerror is not shown.
 (b) The extracted values of $\alpha_{\pom}$.
The full line is a linear fit to $\alpha_{\pom}$ 
with the form given in Eq.(\ref{alpha(t)}).  The dotted
line is the soft Pomeron trajectory~\protect\cite{pl:b348:213}.  The
inner error bars represent the statistical uncertainty; the outer bars
are the statistical and systematic uncertainties added in quadrature.  }
\label{fig-regge}
\vfill
\end{figure}


%
%       ... that's it
%

\providecommand{\etal}{et al.\xspace}
\providecommand{\coll}{Coll.\xspace}
\catcode`\@=11
\def\@bibitem#1{%
\ifmc@bstsupport
  \mc@iftail{#1}%
    {;\newline\ignorespaces}%
    {\ifmc@first\else.\fi\orig@bibitem{#1}}
  \mc@firstfalse
\else
  \mc@iftail{#1}%
    {\ignorespaces}%
    {\orig@bibitem{#1}}%
\fi}%
\catcode`\@=12
\begin{mcbibliography}{10}

\bibitem{epj:c2:247}
ZEUS \coll, J.~Breitweg \etal,
\newblock Eur.\ Phys.\ J.{} C~2~(1998)~247\relax
\relax
\bibitem{epj:c6:603}
ZEUS \coll, J.~Breitweg \etal,
\newblock Eur.\ Phys.\ J.{} C~6~(1999)~603\relax
\relax
\bibitem{epj:c12:393}
ZEUS \coll, J.~Breitweg \etal,
\newblock Eur.\ Phys.\ J.{} C~12~(2000)~393\relax
\relax
\bibitem{epj:c13:371}
H1 \coll, C.~Adloff \etal,
\newblock Eur.\ Phys.\ J.{} C~13~(2000)~371\relax
\relax
\bibitem{pl:b539:25}
H1 \coll, C.~Adloff \etal,
\newblock Phys.\ Lett.{} B~539~(2002)~25\relax
\relax
\bibitem{zfp:c73:73}
ZEUS \coll, M.~Derrick \etal,
\newblock Z.\ Phys.{} C~73~(1996)~73\relax
\relax
\bibitem{pl:b487:273}
ZEUS \coll, J.~Breitweg \etal,
\newblock Phys.\ Lett.{} B~487~(2000)~273\relax
\relax
\bibitem{pl:b377:259}
ZEUS \coll, M.~Derrick \etal,
\newblock Phys.\ Lett.{} B~377~(1996)~259\relax
\relax
\bibitem{pl:b380:220}
ZEUS \coll, M.~Derrick \etal,
\newblock Phys.\ Lett.{} B~380~(1996)~220\relax
\relax
\bibitem{pl:b483:360}
H1 \coll, C.~Adloff \etal,
\newblock Phys.\ Lett.{} B~483~(2000)~360\relax
\relax
\bibitem{epj:c10:373}
H1 \coll, C.~Adloff \etal,
\newblock Eur.\ Phys.\ J.{} C~10~(1999)~373\relax
\relax
\bibitem{epj:c24:345}
ZEUS \coll, S.~Chekanov \etal,
\newblock Eur.\ Phys.\ J.{} C~24~(2002)~345\relax
\relax
\bibitem{Chekanov:2004mw}
ZEUS \coll, S.~Chekanov \etal,
\newblock Nucl.\ Phys.{} B~695~(2004)~3\relax
\relax
\bibitem{pl:b541:251}
H1 \coll, C.~Adloff \etal,
\newblock Phys.\ Lett.{} B~541~(2002)~251\relax
\relax
\bibitem{pl:b437:432}
ZEUS \coll, J.~Breitweg \etal,
\newblock Phys.\ Lett.{} B~437~(1998)~432\relax
\relax
\bibitem{pl:b483:23}
H1 \coll, C.~Adloff \etal,
\newblock Phys.\ Lett.{} B~483~(2000)~23\relax
\relax
\bibitem{pl:b324.knnz}
B.Z.~Kopeliovich, et al.,
\newblock Phys.\ Lett.{} B~324~(1994)~469\relax
\relax
\bibitem{pr:d50:3134}
S.J.~Brodsky \etal,
\newblock Phys.\ Rev.{} D~50~(1994)~3134\relax
\relax
\bibitem{collins:1977:regge}
P.D.B.~Collins,
\newblock {\em An Introduction to {Regge} Theory and High Energy Physics}.
\newblock Cambridge University Press, 1977\relax
\relax
\bibitem{sovjnp:52:529}
M.G.~Ryskin,
\newblock Sov.\ J.\ Nucl.\ Phys.{} 52~(1990)~529\relax
\relax
\bibitem{zfp:c57:89}
M.G.~Ryskin,
\newblock Z.\ Phys.{} C~57~(1993)~89\relax
\relax
\bibitem{pl:b348:213}
A.~Donnachie and P.V.~Landshoff,
\newblock Phys.\ Lett.{} B~348~(1995)~213\relax
\relax
\bibitem{pr:d10:160}
G.A.~Jaroszkiewicz and P.V.~Landshoff,
\newblock Phys.\ Rev.{} D 10~(1974)~170\relax
\relax
\bibitem{FS04}
J.R.~Forshaw and G.~Shaw,
\newblock Preprint \mbox{hep-ph/0411337}, 2004\relax
\relax
\bibitem{misc:forshaw:private}
J. Forshaw, private communication\relax
\relax
\bibitem{pr:d62:14022}
A.D.~Martin, M.G.~Ryskin and T.~Teubner,
\newblock Phys.\ Rev.{} D~62~(2000)~14022\relax
\relax
\bibitem{ivanov:nikolaev:savin}
I.P.~Ivanov, N.N.~Nikolaev, A.A.~Savin,
\newblock Preprint \mbox{hep-ph/0501034}, 2005\relax
\relax
\bibitem{zeus:1993:bluebook}
ZEUS \coll, U.~Holm~(ed.),
\newblock {\em The {ZEUS} Detector}.
\newblock Status Report (unpublished), DESY (1993),
\newblock available on
  \texttt{http://www-zeus.desy.de/bluebook/bluebook.html}\relax
\relax
\bibitem{nim:a279:290}
N.~Harnew \etal,
\newblock Nucl.\ Instr.\ Meth.{} A~279~(1989)~290\relax
\relax
\bibitem{npps:b32:181}
B.~Foster \etal,
\newblock Nucl.\ Phys.\ Proc.\ Suppl.{} B~32~(1993)~181\relax
\relax
\bibitem{nim:a338:254}
B.~Foster \etal,
\newblock Nucl.\ Instr.\ Meth.{} A~338~(1994)~254\relax
\relax
\bibitem{nim:a309:77}
M.~Derrick \etal,
\newblock Nucl.\ Instr.\ Meth.{} A~309~(1991)~77\relax
\relax
\bibitem{nim:a309:101}
A.~Andresen \etal,
\newblock Nucl.\ Instr.\ Meth.{} A~309~(1991)~101\relax
\relax
\bibitem{nim:a321:356}
A.~Caldwell \etal,
\newblock Nucl.\ Instr.\ Meth.{} A~321~(1992)~356\relax
\relax
\bibitem{nim:a336:23}
A.~Bernstein \etal,
\newblock Nucl.\ Instr.\ Meth.{} A~336~(1993)~23\relax
\relax
\bibitem{nim:a450:235}
A.~Bamberger \etal,
\newblock Nucl.\ Instr.\ Meth.{} A~450~(2000)~235\relax
\relax
\bibitem{nim:a401:63}
A.~Bamberger \etal,
\newblock Nucl.\ Instr.\ Meth.{} A~401~(1997)~63\relax
\relax
\bibitem{nim:a277:176}
A.~Dwurazny \etal,
\newblock Nucl.\ Instr.\ Meth.{} A~277~(1989)~176\relax
\relax
\bibitem{acpp:b32:2025}
J.~Andruszk\'ow \etal,
\newblock Acta Phys.\ Pol.{} B~32~(2001)~2025\relax
\relax
\bibitem{pr:129:1834}
L.N.~Hand,
\newblock Phys.\ Rev.{} 129~(1963)~1834\relax
\relax
\bibitem{thesis:miro}
M.~Helbich.
\newblock Ph.D.\ Thesis, Columbia University, 2004, unpublished\relax
\relax
\bibitem{trigger:1992}
W.H.~Smith, K.~Tokoshuku and L.W.~Wiggers,
\newblock {\em Proc. Computing in High-Energy Physics (CHEP), Annecy, France},
  C.~Verkerk and W.~Wojcik~(eds.), p.~222.
\newblock  (1992).
\newblock Also in preprint \mbox{DESY 92-150B}\relax
\relax
\bibitem{nim:a355:278}
W.H.~Smith \etal,
\newblock Nucl.\ Instr.\ Meth.{} A~355~(1995)~278\relax
\relax
\bibitem{tech:cern-dd-ee-84-1}
R.~Brun et al.,
\newblock {\em {\sc geant3}},
\newblock Technical Report CERN-DD/EE/84-1, CERN, 1987\relax
\relax
\bibitem{thesis:muchorowski:1996}
K.~Muchorowski.
\newblock Ph.D.\ Thesis, Warsaw University, Warsaw, Poland, 1996,
  unpublished\relax
\relax
\bibitem{cpc:69:155-tmp-3cfb28c9}
A.~Kwiatkowski, H.~Spiesberger and H.-J.~M\"ohring,
\newblock Comput.\ Phys.\ Comm.{} 69~(1992)~155.
\newblock Also in {\it Proc.\ Workshop Physics at HERA}, W.~Buchm\"{u}ller and
  G.Ingelman (eds.), DESY, Hamburg, (1991)\relax
\relax
\bibitem{spi:www:heracles}
H.~Spiesberger,
\newblock {\em An Event Generator for $ep$ Interactions at {HERA} Including
  Radiative Processes (Version 4.6)}, 1996,
\newblock available on \texttt{http://www.desy.de/\til
  hspiesb/heracles.html}\relax
\relax
\bibitem{thesis:adamczyk:1999}
L.~Adamczyk.
\newblock Ph.D.\ Thesis, University of Mining and Metallurgy, Cracow, Poland,
  Report \mbox{DESY-THESIS-1999-045}, DESY, 1999\relax
\relax
\bibitem{thesis:kasprzak:1994}
M.~Kasprzak.
\newblock Ph.D.\ Thesis, Warsaw University, Warsaw, Poland, Report \mbox{DESY
  F35D-96-16}, DESY, 1996\relax
\relax
\bibitem{pr:d66:31}
Particle Data Group, K.~Hagiwara \etal,
\newblock Phys.\ Rev.{} D~66~(2002)~31\relax
\relax
\bibitem{epj:c21:443}
ZEUS \coll, S.~Chekanov \etal,
\newblock Eur.\ Phys.\ J.{} C~21~(2001)~443\relax
\relax
\bibitem{prl:22:981}
J.J.~Sakurai,
\newblock Phys.\ Rev.\ Lett.{} 22~(1969)~981\relax
\relax
\bibitem{ZEUS-S}
ZEUS Coll., S. Chekanov et al.,
\newblock Phys.\ Rev.{} D~67~(2003)~12007\relax
\relax
\bibitem{CTEQ6M}
J.~Pumplin, et al.,
\newblock JHEP{} 0207~(2002)~012\relax
\relax
\bibitem{MRST99}
A.D.~Martin, et al.,
\newblock Eur.\ Phys.\ J.{} C~14~(2000)~133\relax
\relax
\bibitem{pr:d60:074012}
J.R.~Forshaw, G.~Kerley and G.~Shaw,
\newblock Phys.\ Rev.{} D~60~(1999)~074012\relax
\relax
\bibitem{np:b61:381}
K.~Schilling and G.~Wolf,
\newblock Nucl.\ Phys.{} B~61~(1973)~381\relax
\relax
\bibitem{barone:predazzi}
V.~Barone and E.~Predazzi,
\newblock {\em High Energy Particle Diffraction},
\newblock in Texts and Monographs in Physics.
\newblock Springer Verlag, Berlin (Germany), 2002\relax
\relax
\end{mcbibliography}
\end{document}